\pdfoutput=1
\documentclass[twoside]{article}

\usepackage{graphicx}
\usepackage{amsmath}
\usepackage{amssymb}
\usepackage{float,color}
\usepackage{amsfonts}
\usepackage{subfigure}
\usepackage[margin=1in]{geometry}

\DeclareMathOperator{\sech}{sech}
\DeclareMathOperator{\rand}{rand}

\title{Weakly Nonlinear Analysis of Vortex Formation in a Dissipative Variant of the
Gross-Pitaevskii Equation}

\author{
J.C.~Tzou%
{\thanks{%
Department of Mathematics and Statistics,
Dalhousie University, Halifax, Canada
}},
P.G.~Kevrekidis%
{\thanks{%
Department of Mathematics and Statistics,
University of Massachusetts, Amherst, MA 01003-4515, USA;
Center for Nonlinear Studies and Theoretical Division,
Los Alamos National Laboratory, Los Alamos, NM 87544 USA
}},
T.~Kolokolnikov%
{\thanks{%
Department of Mathematics and Statistics,
Dalhousie University, Halifax, Canada
}},
and
R.~Carretero-Gonz{\'a}lez%
{\thanks{%
Nonlinear Dynamical System Group,
({\texttt{URL}: http://nlds.sdsu.edu}),
Computational Science Research Center,
({\texttt{URL}: http://www.csrc.sdsu.edu}),
and Department of Mathematics and Statistics,
San Diego State University,
San Diego, California 92182-7720, USA
}}
}

\begin{document}
\maketitle
\newcommand{\slugmaster}{%
\slugger{siads}{xxxx}{xx}{x}{x--x}}%

\begin{abstract}
For a dissipative variant of the two-dimensional Gross-Pitaevskii equation with a parabolic trap under rotation, we study a symmetry breaking process that leads to the formation of vortices. The first symmetry breaking leads to the formation of many small vortices distributed uniformly near the Thomas-Fermi radius. The instability occurs as a result of a linear instability of a vortex-free steady state as the rotation is increased above a critical threshold. We focus on the second subsequent symmetry breaking, which occurs in the weakly nonlinear regime. At slightly above threshold, we derive a one-dimensional amplitude equation that describes the slow evolution of the envelope of the initial instability. We show that the mechanism responsible for initiating vortex formation is a modulational instability of the amplitude equation. We also illustrate the role of dissipation in the symmetry breaking process. All analyses are confirmed by detailed numerical computations.
\end{abstract}

\vspace{0.2cm}

\vspace{0.5cm}

\pagestyle{myheadings}
\thispagestyle{plain}
\markboth{J.C.~Tzou, P.G.~Kevrekidis, T.~Kolokolnikov, and R.~Carretero-Gonz{\'a}lez}%
{Vortex Formation in a Dissipative GPE}

\section{Introduction}

The topic of vortex formation upon rotation of an atomic Bose-Einstein
condensate has received a tremendous volume of attention during
the past 15 years, with many of the relevant results finding
their way in main archival references on the subject, including
the books~\cite{becbook1,becbook2}. This is natural, not
only because of the inherent interest in vortices as fundamental
structures in this and more generally in atomic, quantum, and superfluids
systems~\cite{Pismen1999}, but also because this has been a prototypical
way of introducing vortices in the system. These studies not only include
theoretical works but also numerous experiments, in
isotropic and anisotropic settings, with few or with many atoms,
in oblate or prolate traps in at least four distinct experimental
groups pioneering the early experiments~\cite{mit,haljan,ens,oxford}.
Even far more recent experiments, relying chiefly on other
techniques, including the Kibble-Zurek mechanisms utilize rotation
as a way of controllably producing vortices of a given (same)
charge~\cite{hallprl}.

It is then natural to expect a large volume of theoretical
literature tackling the relevant theme. It was realized
early on that the surface excitations play a crucial role
in the relevant ``instability'' that leads to the emergence
of vortices~\cite{dalfovo97}.
The work of Isoshima and Machida~\cite{machida}
was among the first that recognized
the complex energetic balance between the different scenarios
(e.g. stable, metastable or potentially unstable non-vortex states,
and similarly for vortex bearing states). This metastability
opens the potential for hysteretic phenomena, depending on
whether, in accordance with the experiment, the rotation
frequency was ramped up or ramped down, as illustrated, e.g.,
in Ref.~\cite{victor}. Numerous simulations also followed these
earlier works, including, at different levels, finite temperature
considerations. More specifically, both Refs.~\cite{penckwitt2002nucleation},
as well as Ref.~\cite{tsub} considered
the finite temperature model of the so-called dissipative
Gross-Pitaevskii equation (see details below), and of ramps
therein, as a prototypical system where the nucleation and
emergence of vortex lattices was spontaneous. On the
other hand, the work of Ref.~\cite{simula} used the framework
of the Hartree-Fock-Bogoliubov method (in the so-called
Popov approximation) as a means of self-consistently
including thermal effects, finding that the particular
value of the temperature may affect the number of
vortices formed.

From a theoretical perspective, there have been, to the best
of our understanding, two distinct schools of thought.
One of these, based on the work of Ref.~\cite{anglin}
(see also importantly the later interpretation of Ref.~\cite{dubessy},
for the case of a toroidal trap), is based on computing the
Landau criterion threshold, i.e., identifying the order of
the mode that will be associated with the Landau instability
and inferring from that the number of vortices that will
emerge. A distinct approach pioneered by the work of
Stringari and collaborators~\cite{recati,kramer} (see
also Ref.~\cite{sinha}, as well as the review of the relevant
considerations in Ref.~\cite{becbook2}) involved the bifurcation
---from the ground state, be it isotropic or anisotropic--- of
additional states, beyond the rotation frequency that
renders neutral (i.e., of vanishing frequency) the quadrupolar mode.
These two approaches have both been developed in the limit of large
chemical potential, yet to the best of our knowledge,
they have never quite been ``reconciled'' with each other, aside
from a short remark in the work of Ref.~\cite{anglin} suggesting that
Landau method is more relevant when surface excitations are
crucial, while if the instability has a more global character
(e.g., for smaller atom numbers), then the hydrodynamic approach
of Refs.~\cite{becbook2,recati,kramer,sinha} is more suitable.

While understanding these two approaches, their similarities
and differences, and providing a unified perspective of this problem
based on them appears to us an intriguing problem for further
study, we will not pursue it further here. Instead, we will focus
on characterizing exactly how a vortex is ``born'' and migrates
inwards towards the center of the trap.
We will build on our earlier work~\cite{carretero2014vortex} where we
used a multi-scale expansion to obtain a reduction of the
relevant eigenvalue problem, associated with the vortex forming
instability. In our case, where the model of choice is the
dissipative Gross-Pitaevskii equation
(DGPE), we argued that there is a true instability, contrary
to what is the case with the Hamiltonian case, where the
eigenvalues simply cross the origin of the spectral plane
changing ``energy'' or ``signature''
---see the details of Ref.~\cite{carretero2014vortex}.
Here, we take this analysis a significant step further, by reducing
the relevant dynamics, at the periphery of the atomic cloud,
to an effective one-dimensional azimuthal strip.

Remarkably, we find that although the original dynamics pertains
to a {\it self-defocusing} Gross-Pitaevskii equation (GPE), this
reduced azimuthal evolution bears a {\it self-focusing}
character. This trait is manifested through the emergence
of a {\it modulational instability} (MI) against the backdrop
of the homogeneous background. This, in turn, results in a
``spike'' emerging as subtracted from the background, which
finally will morph into a vortex initially rotating along the
strip and gradually spiraling inwards in accordance with its
dynamical equation of motion ---for vortex motion within the
DGPE realm see, e.g., Ref.~\cite{yan2014exploring}\footnote{In the latter
case, the precession corresponds to an anomalous (negative
energy or signature) mode and hence the spiraling occurs
outwards, but in the presence of rotation this mode becomes
normal and similar dynamical equations describe the spiraling
inwards.}.
Our emphasis here will be in highlighting the mechanism
leading to the vortex formation, offering quantitative comparisons
of our focusing GPE reduction (and its MI mechanism) with the
full two-dimensional numerical results. Our presentation
is structured as follows. In Sec.~\ref{sec:math}, we briefly present
the mathematical setup. In Sec.~\ref{sec:weakly}, we discuss the weakly
nonlinear analysis and the derivation of the effective
one-dimensional self-focusing GPE. In Sec.~\ref{sec:MI},
we analyze the MI and compare its predictions to the full system.
Finally in Sec.~\ref{sec:conclu}, we summarize our findings and present
some directions for future study.

\section{Mathematical Setup}
\label{sec:math}

Our starting point will be the dissipative variant of the Gross-Pitaevskii equation (DGPE)
of the form~\cite{pitaevskii1958phenomenological}
(see also the more recent works of Refs.~\cite{penckwitt2002nucleation,tsub})
\begin{equation} \label{GPEdim}
(\gamma - i)u_t = \frac{1}{2}\Delta u + \left(\mu - \frac{1}{2}\Omega_{trap}^2\, \rho^2\right)u - |u|^2u - i\Omega_{\rm rot}u_\theta,
\end{equation}
where $|u(\rho, \theta, t)|^2$ is the time-dependent two-dimensional density of the atomic condensate cloud within a parabolic trap of strength $\Omega_{trap}$ and $\gamma$ accounts for a phenomenological tem\-pe\-ra\-tu\-re-dependent dissipation effect (see, e.g., Refs.~\cite{kavokin2007microcavities, penckwitt2002nucleation, proukakis2008finite, blakie2008dynamics, jackson2009finite, griffin2009bose}). This phenomenological dissipation
term provides a prototypical way of effectively accounting for the interaction of the
condensate with the thermal cloud. Physically relevant values of $\gamma > 0$ are of magnitude $1\times 10^{-3}$; see, e.g., Ref.~\cite{yan2014exploring}.
%
Equation~\eqref{GPEdim} is already written in the co-rotating frame of the trap rotating with frequency $\Omega_{\rm rot}$. The chemical potential $\mu$ is a measure of the strength of interaction between atoms, which we assume to be large in comparison to all other parameters in Eq.~\eqref{GPEdim}. This assumption motivates the following scaling and definitions
$$
t = \frac{1}{\mu}T,
\quad
\rho = \frac{1}{\Omega_{trap}} \sqrt{2\mu}r,
\quad
u(\rho,\theta,t) = \sqrt{\mu} W(r,\theta,T);
$$
where
$$
\widetilde{\Omega} \equiv \frac{1}{\mu}\Omega_{\rm rot},
\quad
\varepsilon \equiv \frac{1}{2\mu}\Omega_{trap} \ll 1,
$$
so that, in rescaled form, the DGPE may be written as
\begin{equation} \label{GPE}
	(\gamma - i)\,W_T = \varepsilon^2\Delta W + (1-r^2)\,W - |W|^2W - i\widetilde{\Omega}W_\theta; \quad \gamma > 0,
\end{equation}
where the edge of the atomic cloud, the Thomas-Fermi radius, is now rescaled
to $r=1$.
A radially symmetric, vortex-free, steady state $W = W_0(r)$ of Eq.~\eqref{GPE} exists and satisfies
$$	
W_{0rr} + \frac{1}{r}W_{0r} + (1-r^2)W_0 - |W_0|^2W_0 = 0, \qquad |W_0| \to 0 \enspace \mbox{as} \enspace r \to \infty.
$$
Notice that the steady state profile is identical to the one of the
corresponding Hamiltonian ($\gamma=0$) model.

Let us use here an approach extending our recent considerations
in Ref.~\cite{carretero2014vortex}.
In that work, for increasing rotation frequency $\widetilde{\Omega}$, it was observed that the steady state $W_0$ first loses stability to a spatial mode scaling as $\mathcal{O}(\varepsilon^{-2/3})$. This instability manifests initially in a large number of small vortices distributed uniformly near the Thomas-Fermi radius
$r = 1$. In fact, the relevant surface mode going unstable
is the one placing in the periphery of the system  the number of
vortices filling it by spanning their respective healing lengths.
This behavior was analyzed in Ref.~\cite{carretero2014vortex}, showing it was due to a linear instability of $W_0$ ---within the DPGE setting, although
the Hamiltonian solution was still identified as dynamically stable---
with increasing rotation frequency. Numerical solutions of Eq.~\eqref{GPE} revealed that a subsequent symmetry breaking mechanism causes only a fraction of these vortices to persist and be pulled into the bulk of the condensate. For our current considerations, for $\widetilde{\Omega}$ slightly above threshold, we perform a weakly nonlinear analysis to examine the onset of this second symmetry breaking process. While the analysis does not predict the fraction of vortices that survive or their eventual fate as they form in the fully nonlinear regime, our analysis accurately captures all of the dynamics in the early stages of their development.
In particular, we show that the weakly nonlinear dynamics of the two-dimensional
{\it self-defocusing} system \eqref{GPE} is described by a one-dimensional perturbed
{\it self-focusing} nonlinear Schr\"{o}dinger equation (NLSE).
This is both perhaps intuitively unexpected and at the same time
crucially relevant to the observed phenomenology.
This is because, as our analysis shows,
the initial pattern selection mechanism responsible for the formation of
vortices is a MI of a non-stationary uniform
solution of the one-dimensional amplitude equation, a mechanism (within
the continuum, cubic nonlinearity considered herein) restricted to the
self-focusing variant of the GPE problem.

\section{Weakly nonlinear analysis and amplitude equations}
\label{sec:weakly}

Since vortices nucleate near the Thomas-Fermi ($r = 1$) radius
with critical wavenumber $m \sim \varepsilon^{-2/3} m_0$ when $\widetilde{\Omega} \sim \varepsilon^{4/3} \Omega$ with $m_0, \Omega \sim \mathcal{O}(1)$, we rescale
Eq.~\eqref{GPE} according to
$$
r = 1+\varepsilon^{2/3}x, \quad
\theta = \varepsilon^{2/3} y, \quad
T = \varepsilon^{-2/3}t, \quad
W = \varepsilon^{1/3}w, \quad
\widetilde{\Omega} = \varepsilon^{4/3}\Omega.
$$
This way, we are restricting our consideration to the small
strip of space near the Thomas-Fermi radius, while considering small amplitude
solutions (since the density approaches zero near that limit), for
longer time scales such that the vorticity is expected to emerge.
In these rescaled variables, Eq.~\eqref{GPE} to leading order becomes
\begin{equation} \label{GPEr}
	(\gamma - i)\,w_t = w_{xx} + w_{yy} - (2x + |w|^2)\,w - i\Omega w_y;
\end{equation}
with
$$
|w| \sim \sqrt{-2x} \enspace \mbox{as} \enspace x \to -\infty,
\quad {\rm and}\quad
|w| \to 0 \enspace \mbox{as} \enspace x \to \infty,
$$
where $w$ is periodic in $y$. In arriving at Eq.~\eqref{GPEr} from Eq.~\eqref{GPE}, the largest terms that have been dropped are of order $\mathcal{O}(\varepsilon^{2/3})$. The focus of the analysis and computations herein will be on Eq.~\eqref{GPEr}.

\begin{figure}[htbp]
  \begin{center}
    \mbox{
    \subfigure[spatial scale] 
        {\label{omegam}
        \includegraphics[width=.45\textwidth]{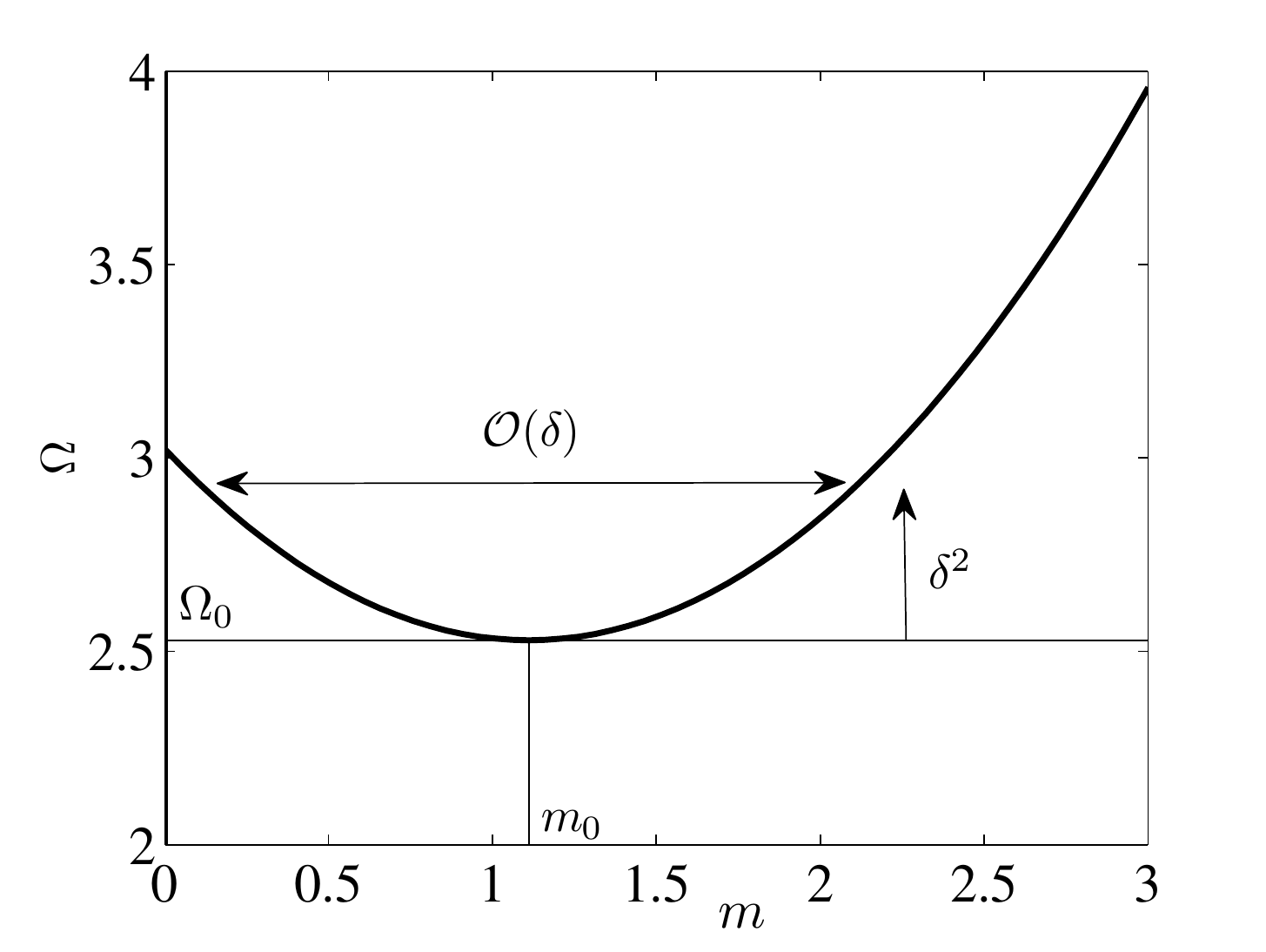}
        }   \hspace{0.15cm}
    \subfigure[eigenvalue scaling] 
        {\label{evs}
        \includegraphics[width=.45\textwidth]{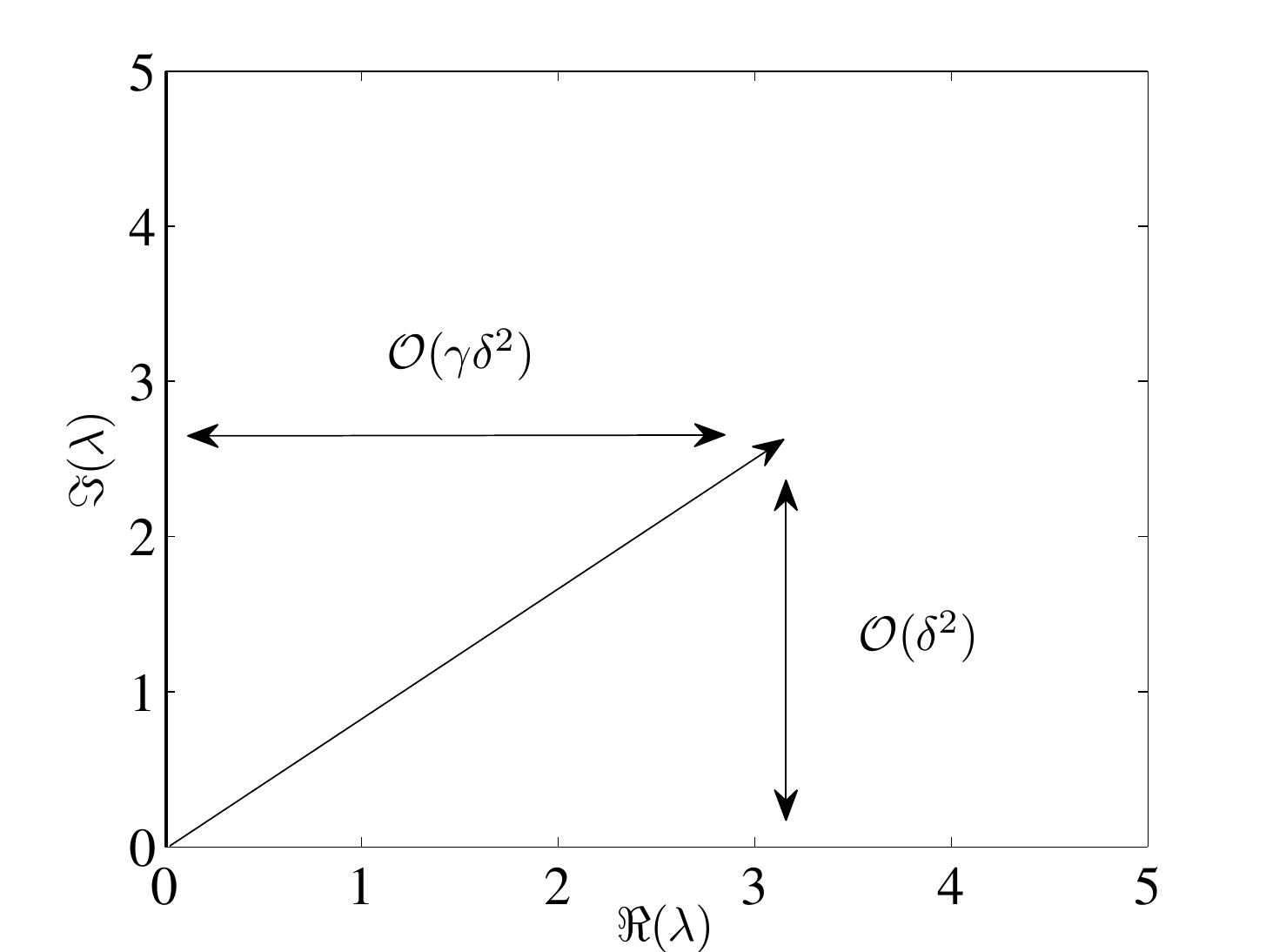}
        }}
    \caption{(a) A depiction of the neutral stability curve $\Omega(m)$ (solid thick line). As $\Omega$ is increased by $\mathcal{O}(\delta^2)$ above $\Omega_0$, an $\mathcal{O}(\delta)$ band of wavenumbers around $m = m_0$ acquires positive growth rate.
    (b) The scalings of $\Re(\lambda)$ and $\Im(\lambda)$ are depicted as $\Omega$ is increased by $\mathcal{O}(\delta^2)$ above $\Omega_0$. The figures here are for illustrative purposes only. See Ref.~\cite{carretero2014vortex} for full results and a detailed analysis.}
  \end{center}
\end{figure}

Writing $w = u + iv$ with $u, v \in \mathbb{R}$, we rewrite Eq.~\eqref{GPEr} as the system
\begin{subequations} \label{realsys}
\begin{eqnarray}
	\gamma u_t + v_t &=& u_{xx} + v_{yy} - (2x + u^2 + v^2)\,u + \Omega v_y,\\[2.0ex]
	\gamma v_t - u_t &=& v_{xx} + v_{yy} - (2x + u^2 + v^2)\,v - \Omega u_y.
\end{eqnarray}
\end{subequations}
A steady state of Eqs.~\eqref{realsys} may be written as $u = u_0(x)$ and $v = 0$,
where $u_0(x)$ is the unique solution of a Painlev\'{e} II equation
$$
	u_0^{\prime\prime} = 2xu_0 + u_0^3,
$$
with limiting conditions (see, e.g., Ref.~\cite{ablowitz1991solitons})
$$
 u_0 \sim \sqrt{-2x} \enspace \mbox{as} \enspace x \to -\infty, \quad
 u_0 \to 0 \enspace \mbox{as} \enspace x \to \infty.
$$
With respect to the full system \eqref{GPE}, $u_0$ is the corner layer near $r = 1$ of the steady state solution $W_0(r)$. Next, we let $u = u_0(x) + \phi(x,y,t)$ and $v = \psi(x,y,t)$
in Eq.~\eqref{realsys} to obtain
\begin{subequations} \label{phipsi}
\begin{eqnarray}
	\gamma \phi_t + \psi_t &=& \phi_{xx} + \phi_{yy} - (2x + 3u_0^2)\,\phi - 3u_0\phi^2 - u_0\psi^2 - \psi^2\phi - \phi^3 + \Omega\psi_y,
\\[2.0ex]
	\gamma \psi_t - \phi_t &=& \psi_{xx} + \psi_{yy} - (2x + u_0^2)\,\psi - 2u_0\phi\psi - \phi^2\psi - \psi^3 - \Omega\phi_y.
\end{eqnarray}
\end{subequations}
The steady state of Eq.~\eqref{phipsi} is then $\phi = \psi = 0$. Assuming a perturbation of the form $(\phi, \psi) = (iA(x), B(x))\,e^{imy+\lambda t}$ in Eq.~\eqref{phipsi}
(given the invariance of the solution along the angular
variable and the periodicity of the latter, we decompose it in Fourier modes)
and collecting linear terms, we obtain the eigenvalue problem
\begin{subequations} \label{eveprob}
\begin{eqnarray}
	A^{\prime\prime} - m^2A - (2x + 3u_0^2)A + m\Omega B_1 &=& \lambda(\gamma A + B),\\[2.0ex]
	B^{\prime\prime} - m^2B -   (2x+ u_0^2)B + m\Omega A_1 &=& \lambda(\gamma B - A).
\end{eqnarray}
\end{subequations}
As in Ref.~\cite{carretero2014vortex}, we set $\lambda = 0$ in Eq.~\eqref{eveprob} and solve the associated eigenvalue problem for $\Omega(m)$, yielding the neutral stability curve depicted in Fig.~\ref{omegam} (see Ref.~\cite{carretero2014vortex} for a detailed analysis and full results). We denote $\Omega_0$ as the smallest value of $\Omega$ at which the steady state loses stability to a perturbation with critical wavenumber $m_0$ (see Fig.~\ref{omegam}). Then, when $\Omega = \Omega_0 + \delta^2$ with $\delta \ll 1$, numerical computations show that $\Re(\lambda) \sim \Im(\lambda)/\gamma \sim \mathcal{O}(\delta^2)$.
This is depicted in Fig.~\ref{evs}.

To analyze the slow evolution of this perturbation slightly above threshold, we assume the asymptotic expansion
\begin{subequations} \label{asymexp}
\begin{equation} \label{vars}
	\Omega = \Omega_0 + \delta^2\Omega_2; \quad
    \phi = \delta\phi_1 + \delta^2\phi_2 + \delta^3\phi_3, \quad
    \psi = \delta\psi_1 + \delta^2\psi_2 + \delta^3\psi_3; \quad
    0< \delta \ll 1.
\end{equation}
The expansion in Eq.~\eqref{vars} is motivated by the expectation that the bifurcation is of the pitchfork type, for which $\phi,\psi \sim \mathcal{O}(\sqrt{\Omega-\Omega_0})$. The solvability condition is then expected to arise at $\mathcal{O}(\delta^3)$. Recalling that the lowest order term omitted from the leading order in Eq.~\eqref{GPEr} is of order $\mathcal{O}(\varepsilon^{2/3})$, we require that $\delta^3 \gg \varepsilon^{2/3}$. We next introduce the slow spatial and temporal scales
\begin{equation} \label{scales}
	x = X/\delta, \quad y = Y/\delta, \quad t = T/\delta^2.
\end{equation}
\end{subequations}
The spatial scale is motivated by the $\mathcal{O}(\delta)$ band of wavenumbers that acquires a positive growth rate as $\Omega$ is increased an $\mathcal{O}(\delta^2)$ distance above threshold (see Fig.~\ref{omegam}), while the temporal scale is motivated by the corresponding scaling of $\lambda$ in Eq.~\eqref{eveprob} (see Fig.~\ref{evs}).
Below, we assume that $\gamma = \mathcal{O}(1)$ with respect to $\delta$.

Substituting Eq.~\eqref{asymexp} into Eq.~\eqref{phipsi}, we solve the linear problems at successively higher orders of $\delta$. At $\mathcal{O}(\delta)$, we obtain the linear
terms of Eq.~\eqref{phipsi}
\begin{subequations} \label{O1}
\begin{eqnarray} \label{O1phi}
\gamma \phi_{1t} + \psi_{1t} &=&
\phi_{1xx} + \phi_{1yy} - (2x + 3u_0^2)\phi_1 + \Omega_0 \psi_{1y},
\\[2.0ex]
\label{O1psi}
\gamma \psi_{1t} - \phi_{1t} &=&
\psi_{1xx} + \psi_{1yy} - (2x + u_0^2)\psi_1 - \Omega_0 \phi_{1y}.
\end{eqnarray}
\end{subequations}
We calculate a $t$-independent solution to Eq.~\eqref{O1} of the form
\begin{equation} \label{O1sol}
\left(\!\begin{array}{c} \phi_1 \\ \psi_1\\ \end{array} \!\right) = C(X,Y,T) \left(\!\begin{array}{c} iA_1(x) \\ B_1(x)\\ \end{array} \!\right)\,e^{im_0y} + c.c.,
\end{equation}
where $c.c.$ denotes the complex conjugate. Here, $A_1(x)$ and $B_1(x)$ are real and satisfy
\begin{subequations}
\begin{equation} \label{A1B1}
L_{m_0} \left(\!\begin{array}{c} A_1 \\ B_1\\ \end{array} \!\right) \equiv \left(\!\begin{array}{c} A_1^{\prime\prime} - m_0^2A_1 - (2x + 3u_0^2)A_1 + m_0\Omega_0B_1
\\[1ex]
B_1^{\prime\prime} - m_0^2B_1 - (2x+ u_0^2)B_1 + m_0\Omega_0A_1\\ \end{array} \!\right) = 0, \quad A_1, B_1 \to 0 \enspace \mbox{as} \enspace x \to \pm \infty,
\end{equation}
with
\begin{equation} \label{m0Omega0}
	m_0 \approx 1.111, \quad \Omega_0 \approx 2.529.
\end{equation}
\end{subequations}
In addition, we impose the normalization constraint
\begin{equation} \label{normalize}
	{\int_{-\infty}^{\infty} \!} A_1^2 + B_1^2 \, dx = 1; \qquad A_1, B_1 > 0.
\end{equation}
In Eq.~\eqref{O1sol}, $C(X,Y,T)$ is a complex quantity that describes the slowly modulated envelope of the perturbation, while in Eq.~\eqref{m0Omega0}, $m_0$ is the critical wavenumber that first becomes unstable as $\Omega$ is increased above $\Omega_0$.

\begin{figure}[htbp]
  \begin{center}
    \mbox{
    \subfigure[$A_1$ and $B_1$] 
        {
        \includegraphics[width=.4\textwidth]{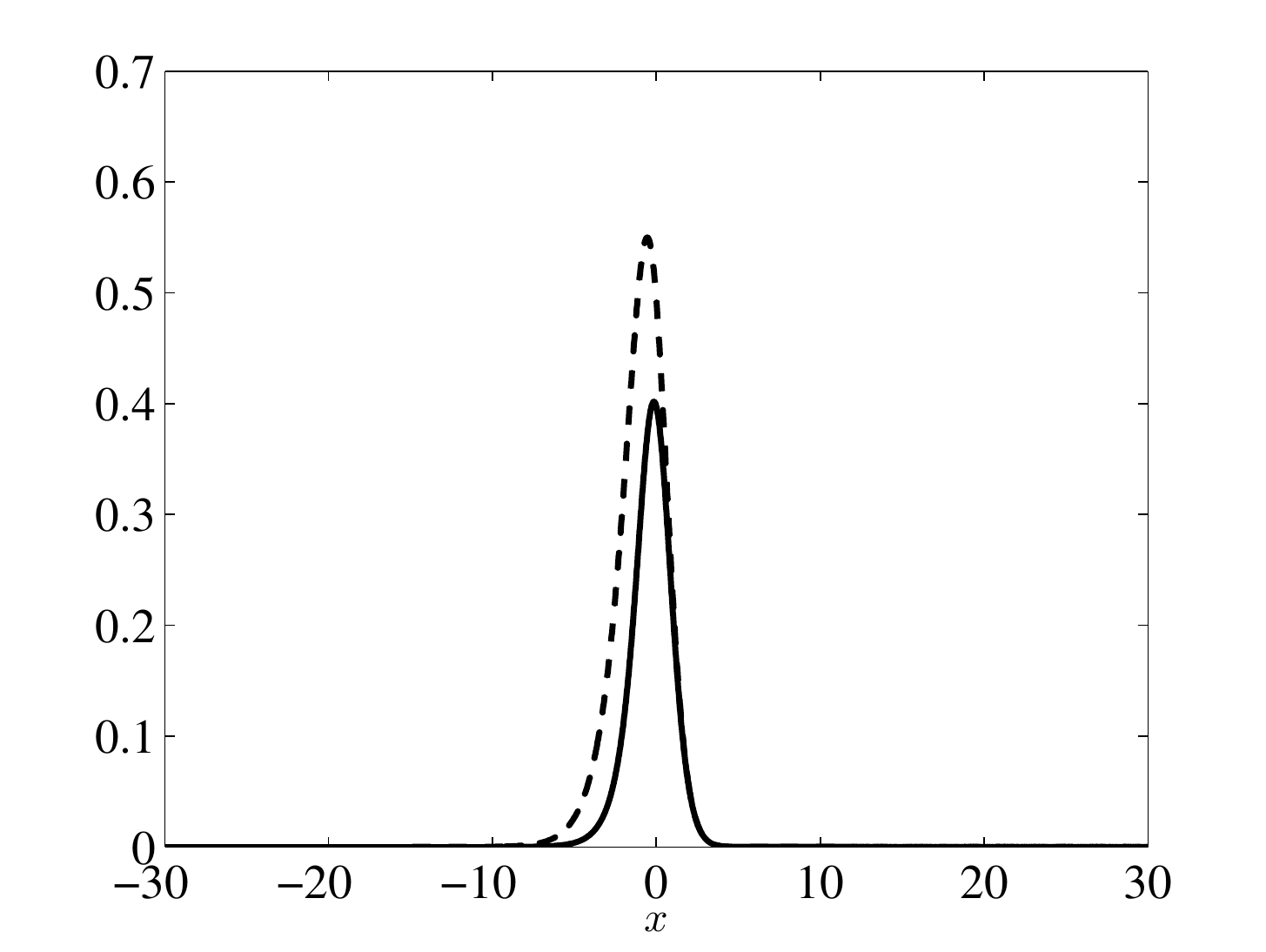}
        }   \hspace{0.15cm}
    \subfigure[$A_{22}$ and $B_{22}$] 
        {
        \includegraphics[width=.4\textwidth]{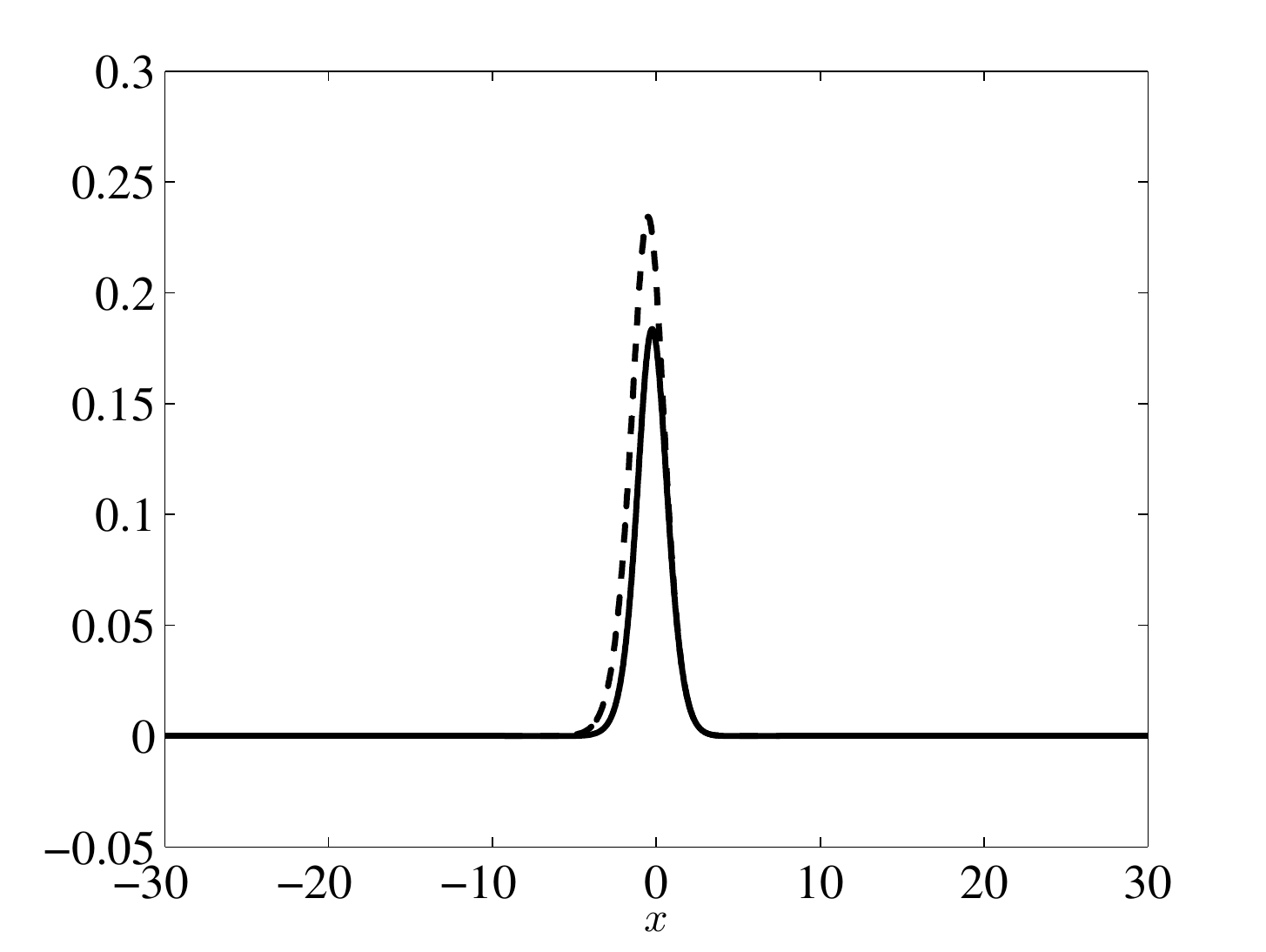}
        }
        }
    \mbox{
    \subfigure[$A_{21}$ and $B_{21}$] 
        {
        \includegraphics[width=.4\textwidth]{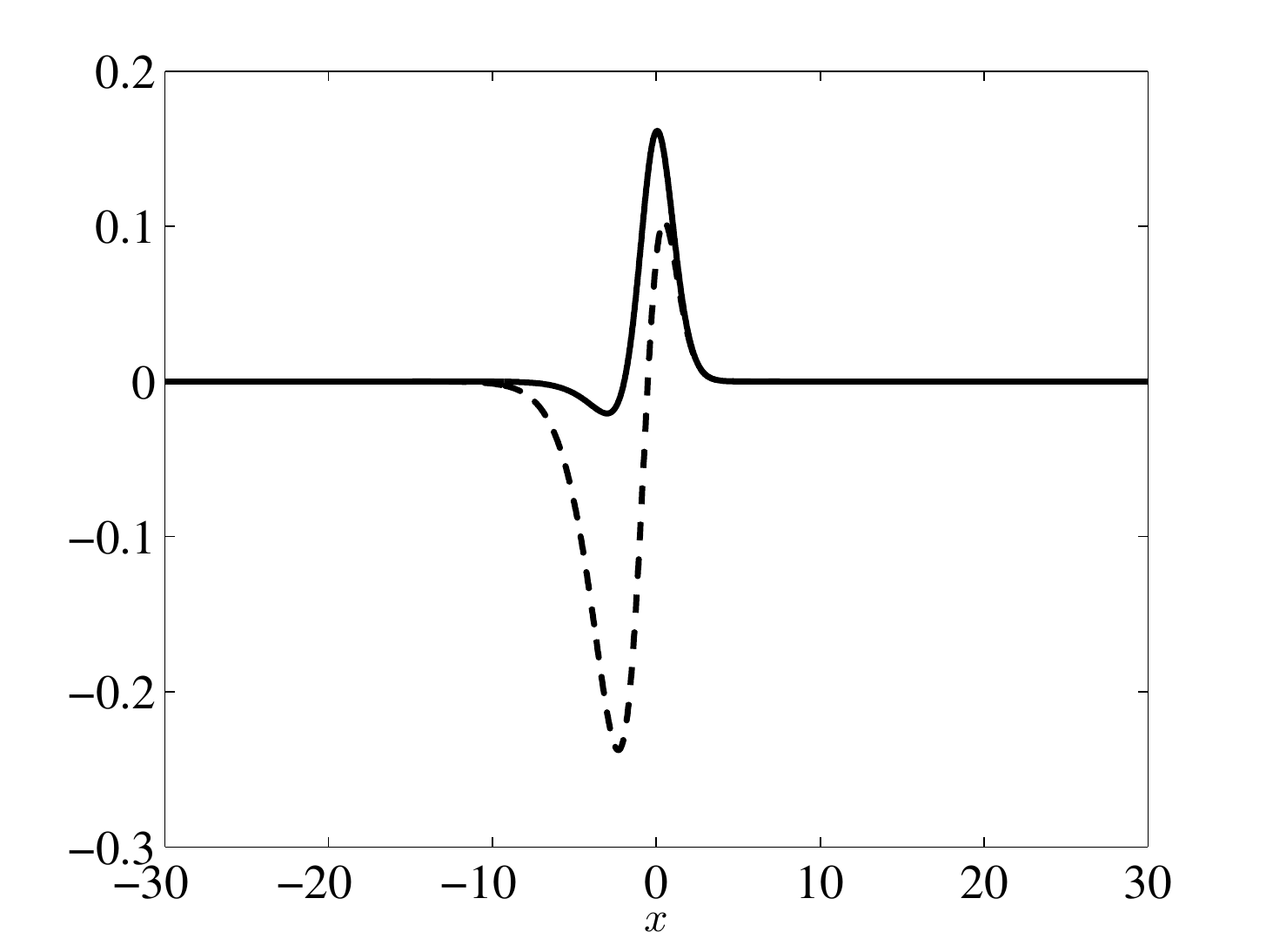}
        }
    \subfigure[$\alpha_{21}$ and $\beta_{21}$] 
        {
        \includegraphics[width=.4\textwidth]{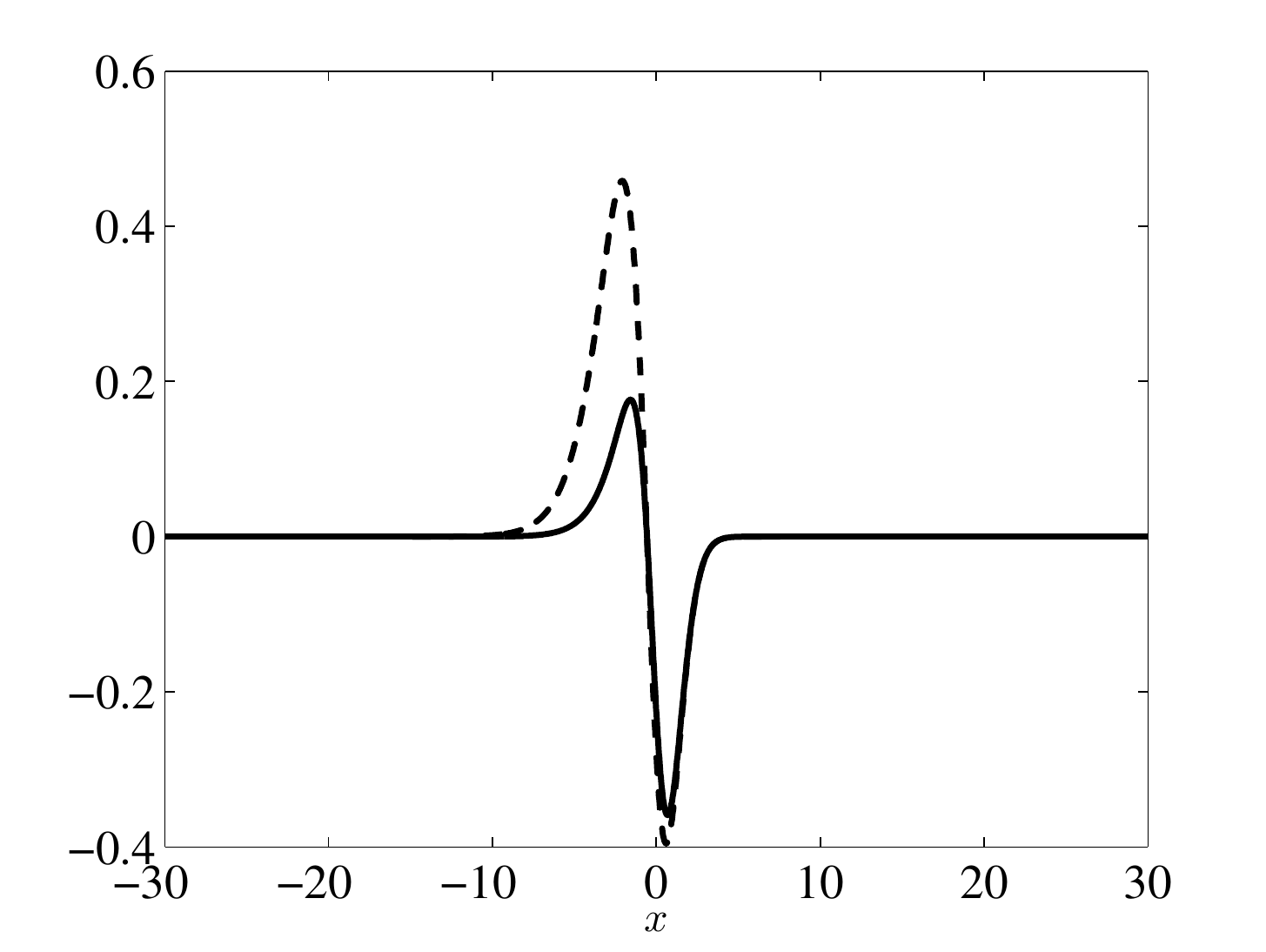}
        }   \hspace{0.15cm}
        }
    \mbox{
    \subfigure[$A_{20}$ and $B_{20}$] 
        {
        \includegraphics[width=.4\textwidth]{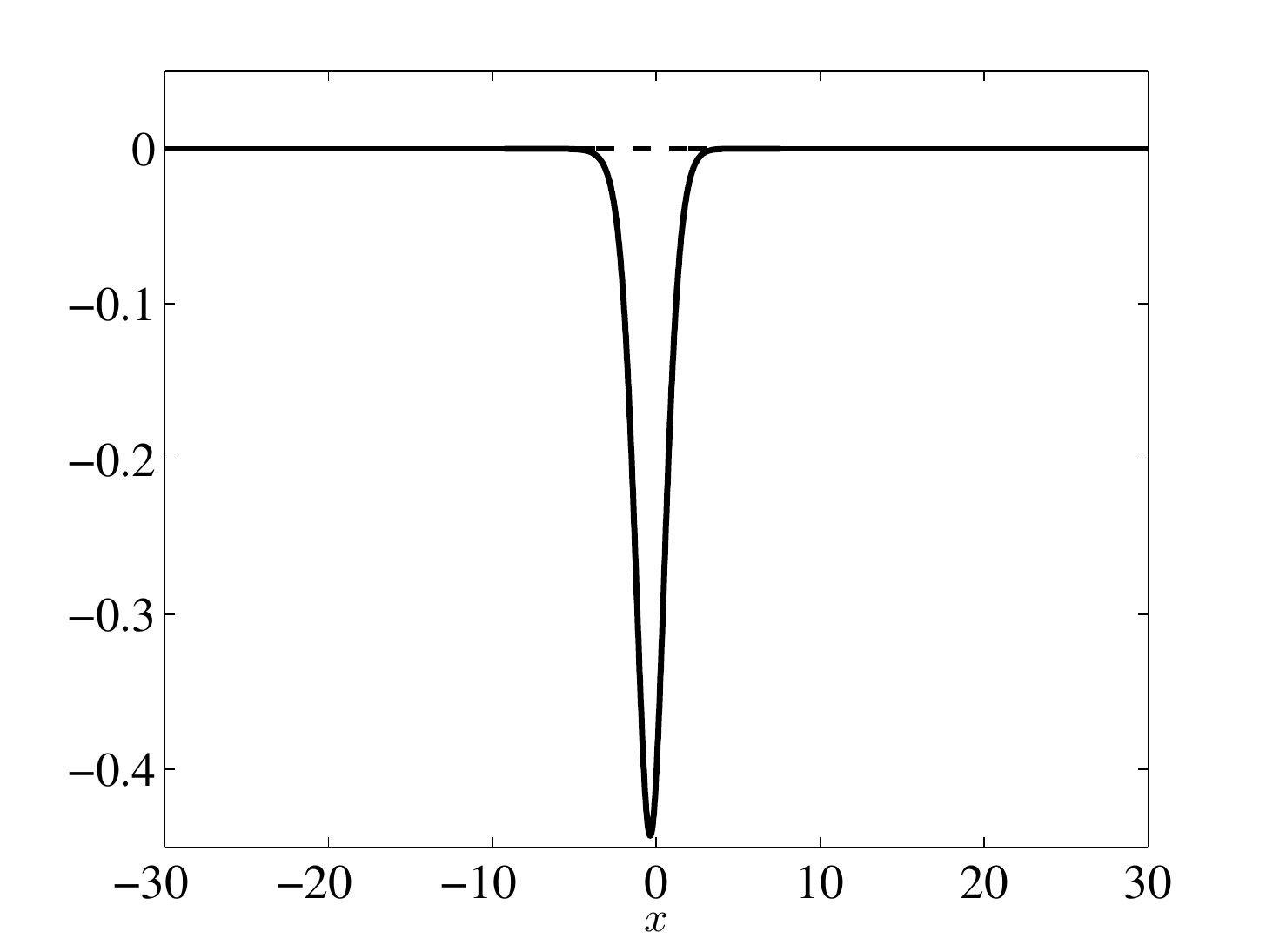}
        }
        }
    \caption{Solutions of Eqs.~\eqref{A22B22}--\eqref{A20B20}, with $A_1$, $A_{22}$, $A_{21}$, $\alpha_{21}$ and $A_{20}$ shown in solid, and $B_1$, $B_{22}$, $B_{21}$, $\beta_{21}$ and $B_{20}$ in dashed.} \label{allsols}
  \end{center}
\end{figure}

At $\mathcal{O}(\delta^2)$, we have
\begin{subequations} \label{O2}
\begin{eqnarray} \label{O2phi}
\qquad
\phi_{2xx} + \phi_{2yy} - (2x + 3u_0^2)\,\phi_2 + \Omega_0 \psi_{2y} &=&
3u_0\phi_1^2 + u_0\psi_1^2 - 2\phi_{1xX} - 2\phi_{1yY} - \Omega_0\psi_{1Y},
\\[2.0ex]
\qquad
\label{O2psi}
\psi_{2xx} + \psi_{2yy} - (2x + u_0^2)\,\psi_2 - \Omega_0 \phi_{2y} &=&
2u_0\phi_1\psi_1 - 2\psi_{1xX} - 2\psi_{1yY} + \Omega_0\psi_{1Y},
\end{eqnarray}
\end{subequations}
with $\phi_1$ and $\psi_1$ given in Eq.~\eqref{O1sol}, the terms on the right-hand sides of Eq.~\eqref{O2} involve terms proportional to $C^2e^{i2m_0y}$, $C_X e^{im_0y}$, $C_Y e^{im_0y}$, and $|C|^2$, along with the corresponding complex conjugates. We therefore write the solution to Eq.~\eqref{O2} as
\begin{multline}
\notag
\left(\!\begin{array}{c} \phi_2 \\ \psi_2\\ \end{array} \!\right) = C^2\left(\!\begin{array}{c} A_{22}(x) \\ -iB_{22}(x)\\ \end{array} \!\right)\,e^{i2m_0y} + \left\lbrack C_Y\left(\!\begin{array}{c} A_{21}(x) \\ -iB_{21}(x)\\ \end{array} \!\right) + C_X\left(\!\begin{array}{c} i\alpha_{21}(x) \\ \beta_{21}(x)\\ \end{array} \!\right)\right\rbrack\,e^{im_0y} + \\ +  |C|^2\left(\!\begin{array}{c} A_{20} \\ B_{20}(x)\\ \end{array} \!\right) + c.c.,
\end{multline}
where the equations for $A_{22}(x)$, $B_{22}(x), \ldots$, are given by
\begin{equation} \label{A22B22}
L_{2m_0} \left(\!\begin{array}{c} A_{22} \\ B_{22}\\ \end{array} \!\right) = \left(\!\begin{array}{c} -3u_0A_1^2 + u_0B_1^2\\ -2u_0A_1B_1\\ \end{array} \!\right), \qquad A_{22}, B_{22} \to 0 \enspace \mbox{as} \enspace x \to \pm \infty,
\end{equation}
\begin{subequations} \label{ABalphabeta}
\begin{equation} \label{A21B21}
L_{m_0} \left(\!\begin{array}{c} A_{21} \\ B_{21}\\ \end{array} \!\right) = \left(\!\begin{array}{c} 2m_0A_1 - \Omega_0B_1 \\ 2m_0B_1-\Omega_0A_1\\ \end{array} \!\right), \qquad A_{21}, B_{21} \to 0 \enspace \mbox{as} \enspace x \to \pm \infty,
\end{equation}
\begin{equation} \label{alphabeta}
L_{m_0} \left(\!\begin{array}{c} \alpha_{21} \\ \beta_{21}\\ \end{array} \!\right) = \left(\!\begin{array}{c} -2A_1^\prime \\ -2B_1^\prime \\ \end{array} \!\right), \qquad \alpha_{21}, \beta_{21} \to 0 \enspace \mbox{as} \enspace x \to \pm \infty,
\end{equation}
\end{subequations}
and
\begin{equation} \label{A20B20}
	L_0 \left(\!\begin{array}{c} A_{20} \\ B_{20}\\ \end{array} \!\right) = \left(\!\begin{array}{c} 6u_0A_1^2 + 2u_0B_1^2 \\ 0 \\ \end{array} \!\right), \qquad A_{20}, B_{20} \to 0 \enspace \mbox{as} \enspace x \to \pm \infty.
\end{equation}
In Eqs.~\eqref{A22B22}--\eqref{A20B20}, the linear operator $L_{m_0}$ is defined in Eq.~\eqref{A1B1}. Since there exists a non-trivial solution to the self-adjoint system \eqref{A1B1}, for solutions to Eq.~\eqref{ABalphabeta} to exist, the right-hand sides must each satisfy the Fredholm conditions
\begin{equation} \label{O2fred}
	{\int_{-\infty}^{\infty} \!} (A_1, B_1)\left(\!\begin{array}{c} 2m_0A_1 - \Omega_0B_1 \\ 2m_0B_1-\Omega_0A_1\\ \end{array} \!\right) \, dx = 0
 \quad {\rm and}\quad
  {\int_{-\infty}^{\infty} \!} (A_1, B_1)\left(\!\begin{array}{c} -2A_1^\prime \\ -2B_1^\prime \\ \end{array} \!\right) \, dx = 0.
\end{equation}
The second condition in Eq.~\eqref{O2fred} may be seen from integrating by parts once and applying the boundary conditions in Eq.~\eqref{A1B1}. The first condition may be inferred from the fact that a solution to Eq.~\eqref{A21B21} exists and is given by $(A_{21}, B_{21}) = \partial_{m_0} (A_1, B_1)$, which may be seen by differentiating Eq.~\eqref{A1B1} with respect to $m_0$ while noting that $d\Omega_0/dm_0 = 0$. This condition, along with the the normalization constraint in Eq.~\eqref{normalize}, yields the identity
$$
	{\int_{-\infty}^{\infty} \!} A_1 B_1 \, dx = \frac{m_0}{\Omega_0}.
$$
With Eq.~\eqref{O2fred} satisfied, we impose the additional orthogonality constraints
$$
{\int_{-\infty}^{\infty} \!} (A_1,B_1) \left(\!\begin{array}{c} A_{21} \\ B_{21}\\ \end{array} \!\right) \,dx = 0,
\quad {\rm and} \quad
{\int_{-\infty}^{\infty} \!} (A_1,B_1) \left(\!\begin{array}{c} \alpha_{21} \\ \beta_{21}\\ \end{array} \!\right)\,dx = 0,
$$
to uniquely specify $A_{21}$, $B_{21}$, $\alpha_{21}$ and $\beta_{21}$. The solutions of Eqs.~\eqref{A22B22}--\eqref{A20B20} are depicted in Fig.~\ref{allsols}.

At $\mathcal{O}(\delta^3)$, we have that
\begin{subequations} \label{O3}
\begin{eqnarray} \label{O3phi}
\phi_{3xx} + \phi_{3yy} - (2x + 3u_0^2)\phi_3 + \Omega_0 \psi_{3y} &=& R_\phi,
\\[2.0ex]
 \label{O3psi}
\psi_{3xx} + \psi_{3yy} - (2x + u_0^2)\psi_3 - \Omega_0 \phi_{3y} &=& R_\psi,
\end{eqnarray}
\end{subequations}
where $R_\phi$ and $R_\psi$ contain the secular terms $S_\phi(x)\,e^{im_0y}$ and $S_\psi(x)\,e^{im_0y}$, respectively, along with other non-resonant terms that we need not consider. For completeness, we give the explicit expressions for the amplitudes
of these secular term:
\begin{multline}
\notag
S_\phi = \left( \gamma i A_1 + B_1\right) \frac{\partial C}{\partial T} - i\left(2\alpha_{21}^\prime + A_1 \right)\frac{\partial^2C}{\partial X^2} + \left(2m_0\alpha_{21} - \Omega_0\beta_{21} - 2a_{21}^\prime \right)\frac{\partial^2C}{\partial X \partial Y} + \\ + i\left(\Omega_0b_{21}- 2m_0a_{21} -A_1  \right)\frac{\partial^2C}{\partial Y^2} -  i\Omega_2m_0B_1C + \\ + i\left(2u_0(3A_1a_{20} -3A_1a_{22} - u_0B_1b_{22}) + 3A^3 + A_1B_1^2 \right)|C|^2C,
\\[2.0ex]
\notag
S_\psi = \left( \gamma B_1 - iA_1\right) \frac{\partial C}{\partial T} - \left(2\beta_{21}^\prime + B_1 \right)\frac{\partial^2C}{\partial X^2} - i\left(2m_0\beta_{21} - \Omega_0 \alpha_{21} - 2b_{21}^\prime \right)\frac{\partial^2C}{\partial X \partial Y} + \\ +  \left(\Omega_0a_{21}-2m_0b_{21} -B_1 \right)\frac{\partial^2C}{\partial Y^2} -  \Omega_2m_0A_1C  + \\ + \left\lbrack2u_0\left(B_1a_{22} - A_1b_{22} + B_1a_{20} \right) + A_1^2B_1 + 3B_1^3 \right\rbrack|C|^2C.
\end{multline}
The solution to Eq.~\eqref{O3} may then be written as $(\phi_3, \psi_3) = (iA_{31}(x), B_{31}(x))\,e^{im_0y} + \ldots$, where $A_{31}$ and $B_{31}$ satisfy the system
\begin{equation} \label{A3B3}
	L_{m_0} \left(\!\begin{array}{c} A_{31} \\ B_{31}\\ \end{array} \!\right) = \left(\!\begin{array}{c} -iS_\phi \\ S_\psi \\ \end{array} \!\right).
\end{equation}
Applying the Fredholm condition to Eq.~\eqref{A3B3}
$$
	{\int_{-\infty}^{\infty} \!} (A_1, B_1)\left(\!\begin{array}{c} -iS_\phi \\ S_\psi \\ \end{array} \!\right)\,dx = 0,
$$
we obtain the following amplitude equation for $C(X,Y,T)$:
\begin{equation} \label{ampeq}
	\left(i\tau_1 - \gamma\tau_2\right) \frac{\partial C}{\partial T} + D_{XX} \frac{\partial^2 C}{\partial X^2} + iD_{XY} \frac{\partial^2 C}{\partial X
\partial Y} + D_{YY}\frac{\partial^2 C}{\partial Y^2} + \sigma C + \alpha|C|^2C = 0.
\end{equation}
The coefficients in Eq.~\eqref{ampeq} are all real, and are given by
\begin{eqnarray}
\tau_1 &=& \frac{2m_0}{\Omega_0}  > 0, \quad \tau_2 = 1, \quad
\sigma = \frac{2m_0^2}{\Omega_0}\Omega_2 > 0,
\notag
\\[2.0ex]
\notag
\alpha &=& -{\int_{-\infty}^{\infty} \!} A_1 \lbrack 2u_0(-3A_1A_{22} + 3A_1A_{20} - B_1B_{22}) + 3A_1^2 + A_1B_1^2 \rbrack +
\notag
\\ [.80ex]
\notag
&&B_1\lbrack 2u_0(-A_1B_{22} + A_{20}B_1 + A_{22}B_1) + A_1^2B_1 + 3B_1^3 \rbrack \,dx > 0,
\notag
\\[2.0ex]
\notag
D_{XX} &=& {\int_{-\infty}^{\infty} \!} A_1 \lbrack -2\alpha_{21}^\prime - A_1\rbrack + B_1\lbrack -2\beta_{21}^\prime - B_1\rbrack \,dx \approx 0,
\notag
\\ [2.0ex]
\notag
D_{XY} &=& {\int_{-\infty}^{\infty} \!} A_1 \lbrack 2A_{21}^\prime - 2m_0\alpha_{21} + \Omega_0\beta_{21}\rbrack + B_1\lbrack 2B_{21}^\prime - 2m_0\beta_{21} + \Omega_0\alpha_{21} \rbrack \,dx \approx 0,
\notag
\\ [2.0ex]
\notag
D_{YY} &=& -{\int_{-\infty}^{\infty} \!} A_1 \lbrack -A_1 - 2m_0A_{21} + \Omega_0B_{21} \rbrack + B_1\lbrack -B_1 - 2m_0B_{21} + \Omega_0A_{21} \rbrack \,dx > 0.
\end{eqnarray}
We note that $\sigma > 0$ since we assume that the system is above threshold;
i.e., $\Omega_2 > 0$ so that $\Omega > \Omega_0$. With the normalization \eqref{normalize} and values of $m_0$ and $\Omega_0$ given in Eq.~\eqref{m0Omega0}, we numerically obtain the following values for the coefficients, accurate to the fifth decimal place:
\begin{equation} \label{numericalvalues}
\begin{gathered}
 \tau_1 \approx 0.87884, \quad
 \tau_2 = 1 , \quad
 \sigma \approx 0.97671 \Omega_2, \quad
 \alpha \approx 0.62184, \quad
 \\[2.0ex]
 D_{YY} \approx 0.67615, \quad
 D_{XY} \sim 10^{-12}, \quad {\rm and} \quad
 D_{XX} \sim 10^{-6}.
\end{gathered}
\end{equation}
The values above show that, remarkably, $D_{XX}$ and $D_{XY}$ are very close to zero.
While they may or may not be exactly zero, for
all practical purposes hereafter, we will indeed set them to $0$.
In this way, the two-dimensional dynamics in the weakly nonlinear regime of Eq.~\eqref{GPEr} reduce to dynamics along only {\em one} dimension. The same reduction was observed for the one-dimensional dynamics of edge modes in a two-dimensional NLSE in the presence of a honeycomb potential~\cite{ablowitz2014adiabatic}.
Indeed, this may be an indication (exactly, or just approximately so)
of an effective
``topological protection''~\cite{ablowitz2}, a theme of intense recent interest
in the physics community~\cite{segev}. The reason we indicate
the potentially approximate nature of the topological
protection for our (toroidal) domain strip is that eventually the
ensuing vortices escape inwards towards the center of
the domain. Nevertheless, exploring this aspect in the context of the present
work further is an especially appealing aspect for further study.

Further proceeding with our reduction, by neglecting the
$D_{XX}$ and $D_{XY}$ terms as indicated above,
our analysis yields the one-dimensional amplitude equation for the envelope $C(Y,T)$
\begin{equation} \label{ampeq1d}
	\left(i\tau_1 - \gamma\right) \frac{\partial C}{\partial T} + D_{YY} \frac{\partial^2 C}{\partial Y^2} + \sigma C + \alpha|C|^2C  = 0.
\end{equation}
We make one remark regarding the scaling of $\gamma$ with respect to $\delta$. Due to the $C \to -C$ invariance of Eq.~\eqref{O1sol}, the largest term omitted from the amplitude equation \eqref{ampeq1d} is the quintic term $|C|^4 C$. This term is of $\mathcal{O}(\delta^2)$ with respect to the rest of the terms in Eq.~\eqref{ampeq1d}. Therefore, while we assumed in the analysis that $\gamma = \mathcal{O}(1)$ with respect to $\delta$, we in fact only require that $\gamma \gg \delta^2$ for Eq.~\eqref{ampeq1d} to be valid. Further, as we show in the next section, when $\gamma$ is small, it is responsible only for the growth of low wavenumber perturbations of a spatially uniform state, and the dissipation of high wavenumber perturbations. The symmetry breaking mechanism in Eq.~\eqref{ampeq1d} that initiates vortex formation in the full system \eqref{GPEr} occurs on an $\mathcal{O}(1)$ time scale independent of $\gamma$. As such, Eq.~\eqref{ampeq1d} retains all the orders required to accurately describe the weakly nonlinear dynamics of the full system.

\begin{figure}[htbp]
  \begin{center}
    \mbox{
    \subfigure[$t = 0$] 
        {\label{fig1}
        \includegraphics[width=.45\textwidth]{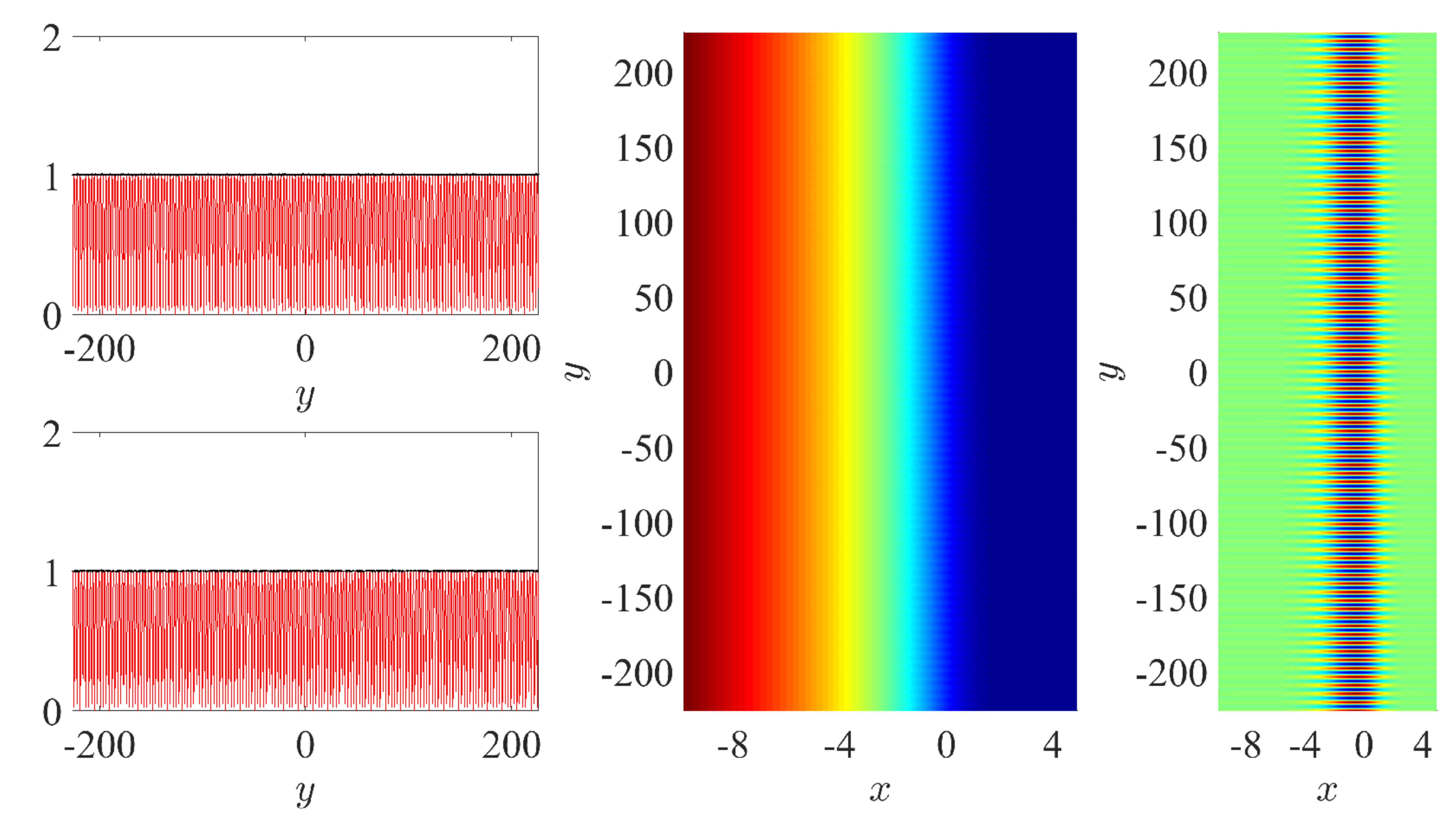}
        }   
    \subfigure[$t = 5343.75$] 
        {\label{fig106}
        \includegraphics[width=.45\textwidth]{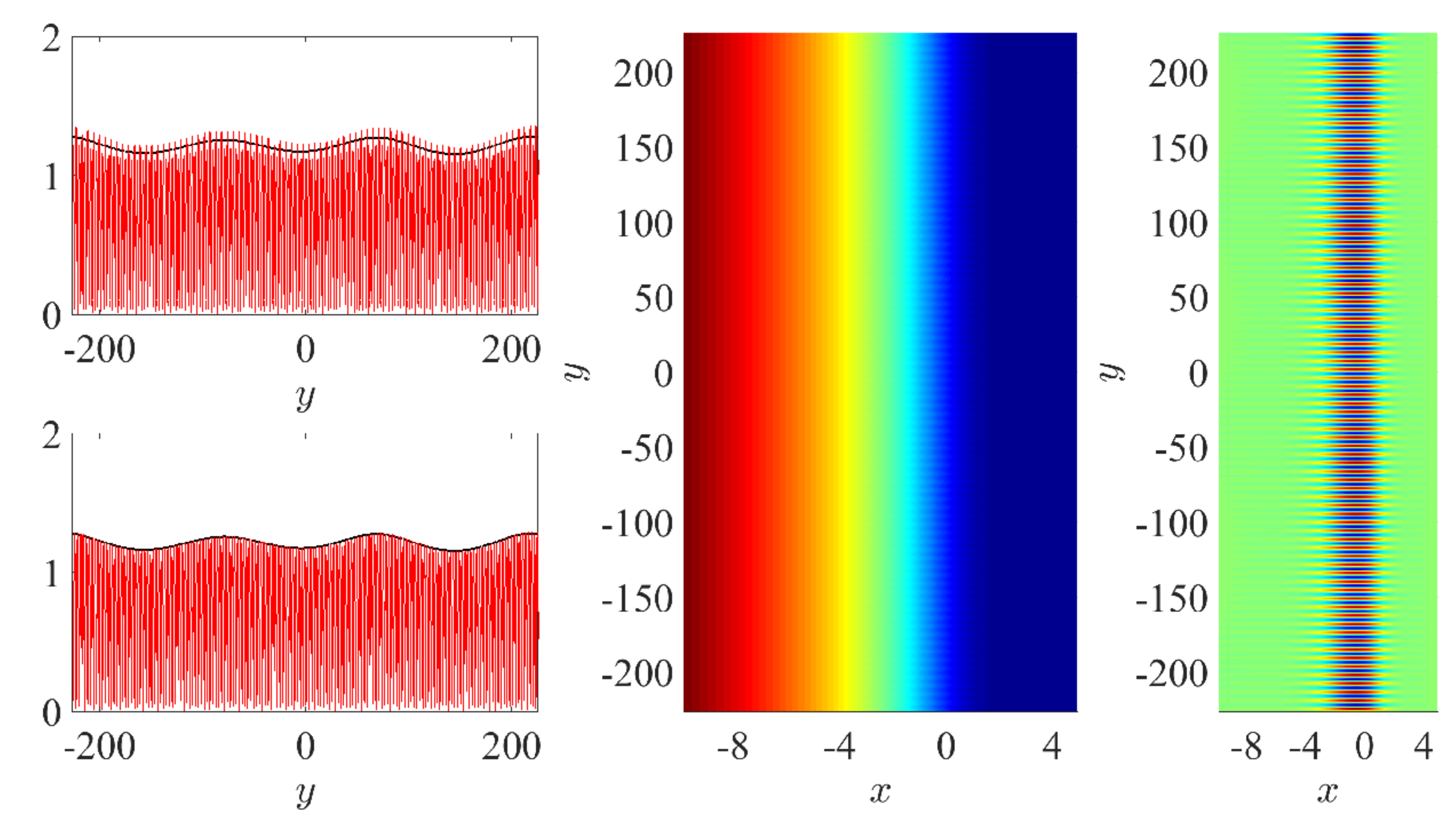}
        }}
    \mbox{
    \subfigure[$t = 6750$] 
        {\label{fig121}
        \includegraphics[width=.45\textwidth]{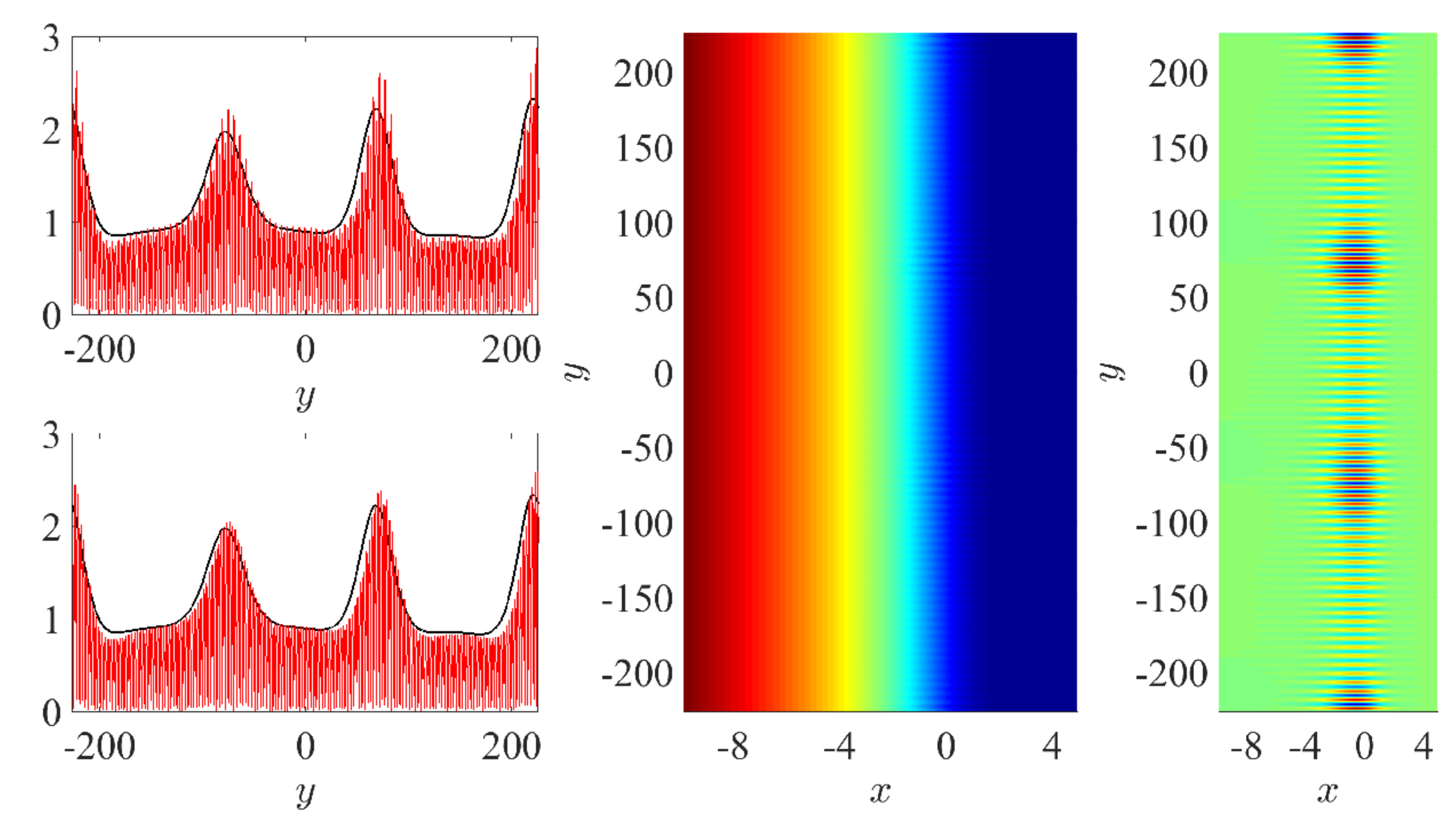}
        }   
    \subfigure[$t = 7171.875$] 
        {\label{fig129}
        \includegraphics[width=.45\textwidth]{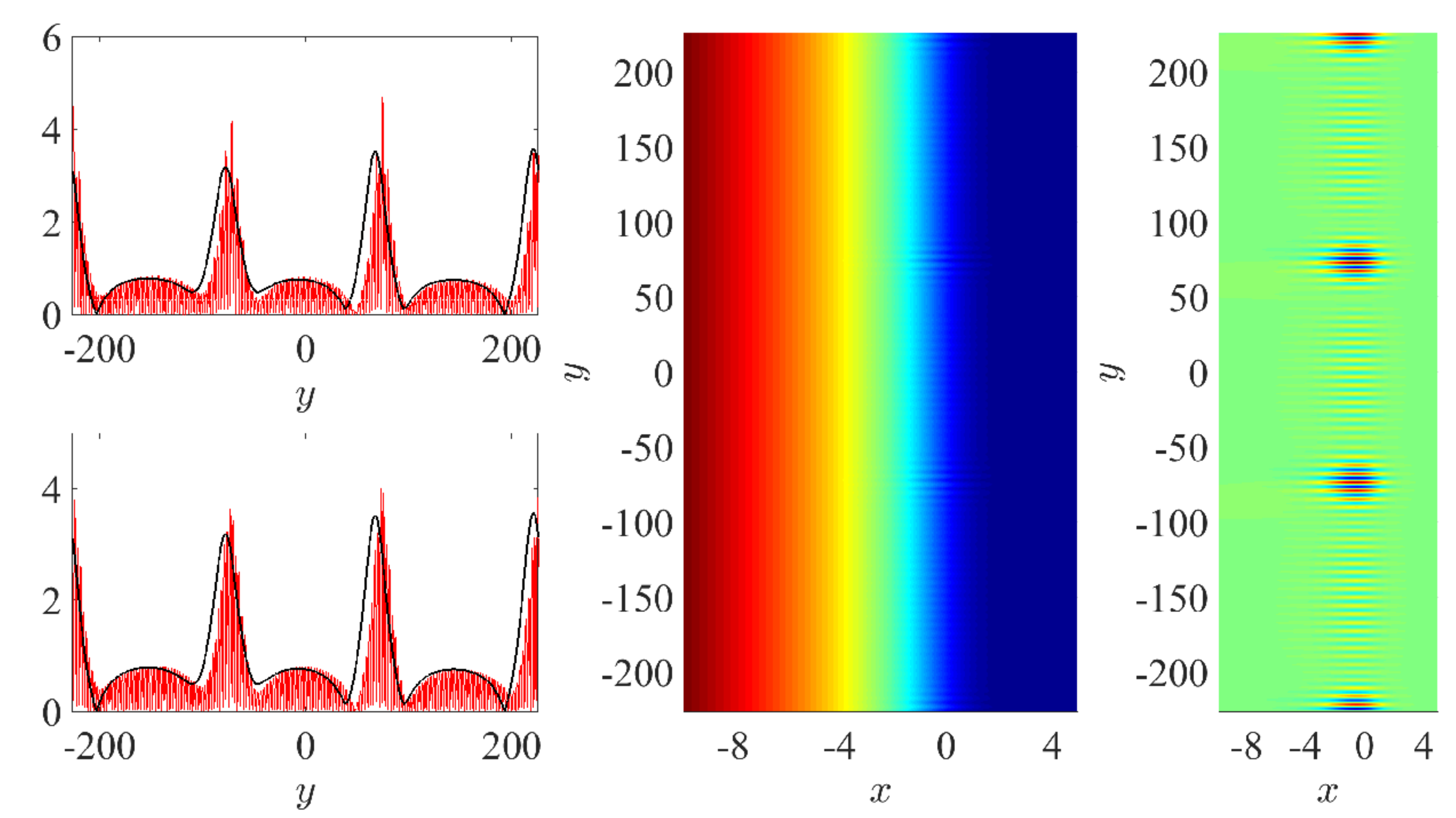}
        }}
    \mbox{
    \subfigure[$t = 8859.375$] 
        {\label{fig64}
        \includegraphics[width=.45\textwidth]{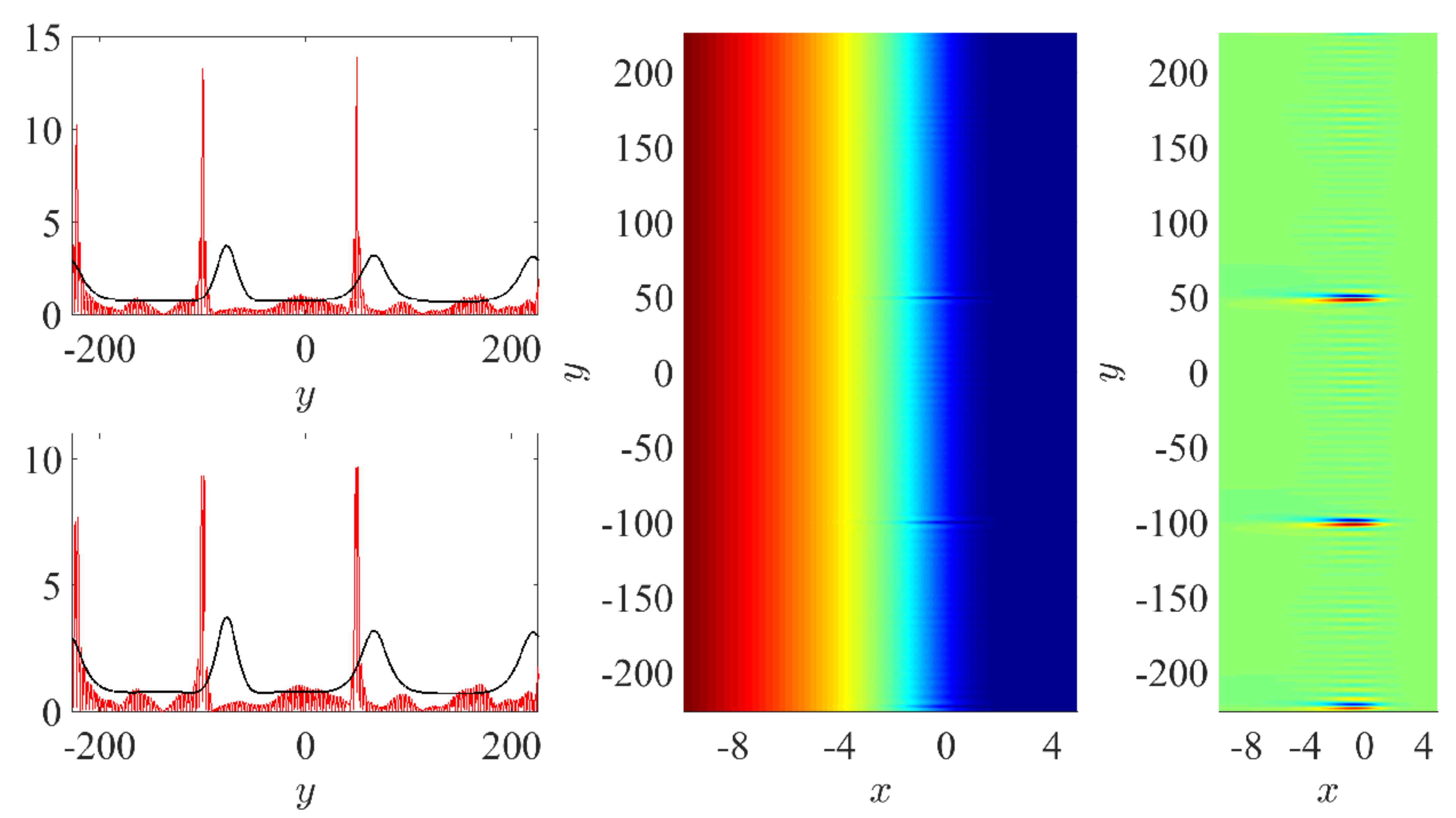}
        }   
    \subfigure[$t = 9421.875$] 
        {\label{fig68}
        \includegraphics[width=.45\textwidth]{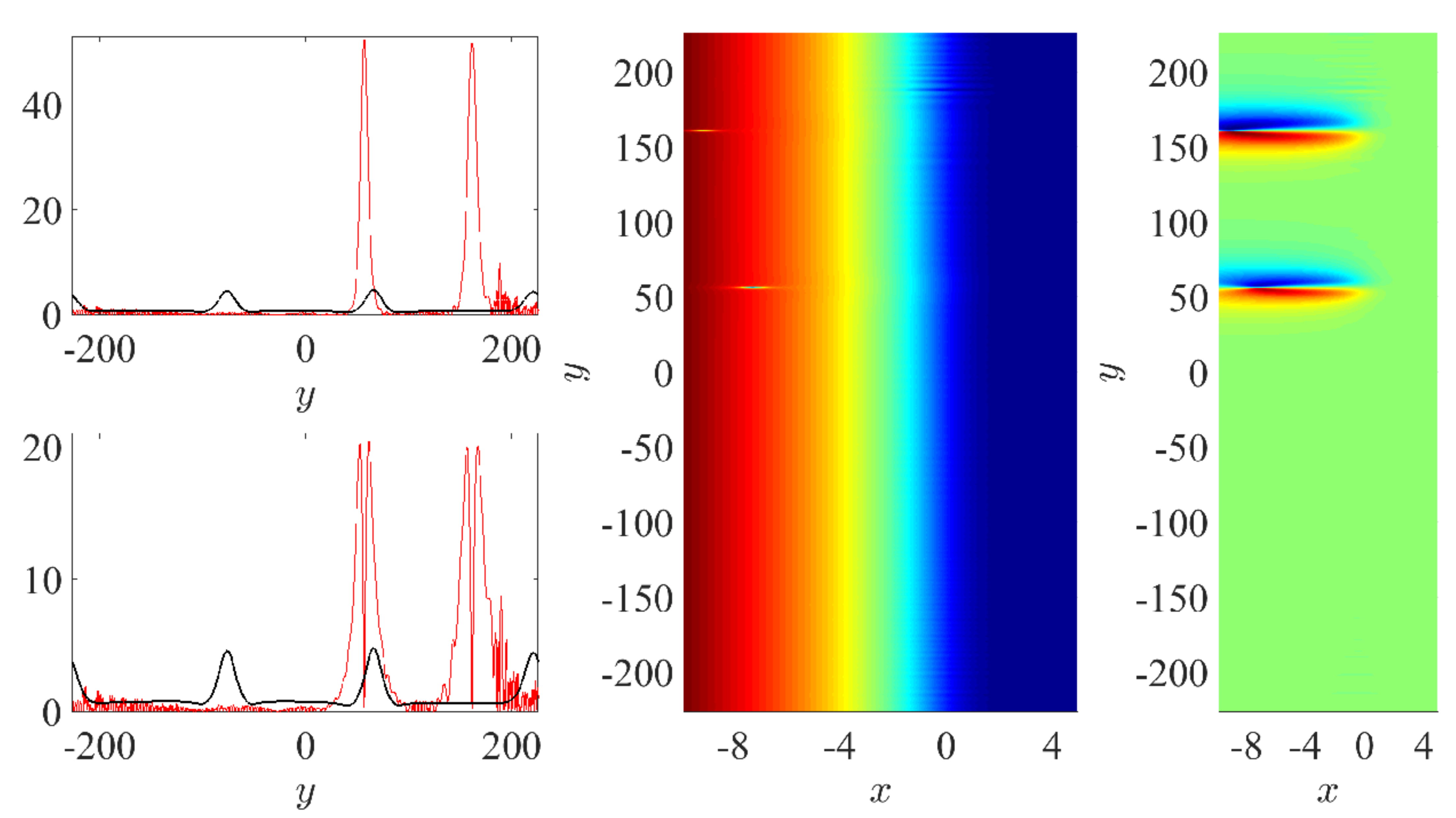}
        }}
    \caption{Evolution of the perturbation at the periphery of the
    atomic cloud. Each panel (a)--(f) is arranged into a left, center, and right column. The center and right columns depict, respectively, the amplitude and imaginary part of the solution, where blue (red) regions indicate small (large) values. The left columns depict the comparison of the one-dimensional dynamics of the amplitude Eq.~\eqref{ampeq1d} (black) versus the dynamics of the full two-dimensional system Eq.~\eqref{GPEr} (red). The red is taken from an $x$-slice of $\phi$ (top subpanel) and $\psi$ (bottom subpanel), the real and imaginary parts of the perturbation, respectively.
    Starting from random perturbations of the envelope about unity (a),
    the third mode is selected (b).
    The pattern oscillates between a slightly more localized state (c)
    and a highly localized state (d), before entering the fully nonlinear
    regime (e) where two vortices are nucleated and pulled into the bulk (f).}
    \label{envelope}
  \end{center}
\end{figure}

To examine the validity of
the weakly nonlinear theory, we solved the two-dimensional system \eqref{GPEr} numerically on the domain $x \in \lbrack -7.5, 22.5\rbrack$, $y \in \lbrack-80\pi/m_0 , 80\pi/m_0\rbrack$ so that exactly $80$ wavelengths of the critical mode $m_0$ fit inside the domain of length $L_y = 160\pi/m_0$. The initial conditions were taken to be of the form given in Eq.~\eqref{O1sol} with an envelope randomly perturbed from unity. That is $C(Y,0) = 1 + 0.01*\rand(y)$, where $\rand(y)$ takes on a uniformly distributed random value between $0$ and $1$ at each discrete point in $y$. The parameters $\gamma$ and $\delta$ were taken to be $\gamma = 0.01$ and $\delta = 0.04$.
While realistic values of $\gamma$ are typically smaller than
$0.01$, this was purely for demonstration purposes and similar results
are obtained for smaller, more realistic, values of $\gamma$.
We also simultaneously solved the one-dimensional amplitude equation \eqref{ampeq1d} on the domain $Y \in \lbrack-80\delta\pi/m_0 , 80\delta\pi/m_0\rbrack$; that is, on a domain of length $L = \delta L_y$, consistent with the scaling in Eq.~\eqref{scales}. The comparison of the two sets of
results is shown in Fig.~\ref{envelope}. Each panel \ref{fig1}--\ref{fig68} is arranged into a left, center, and right column. In the center column of each figure, we show a surface plot of $|w|$, while in the right column, we show a surface plot of $\Im(w) = \psi$. Blue (red) regions indicate small (large) values in the plotted quantity. In the two plots that make up the leftmost column, we show in red a slice of $|\phi|/\delta$ (top) and $|\psi|/\delta$ (bottom) taken near $x = 0$, corresponding to the vicinity of the Thomas-Fermi radius where vortices first form in the original system \eqref{GPE}. Here, $\phi$ and $\psi$ are the real and imaginary parts of the perturbation, respectively, and obey Eq.~\eqref{phipsi}. In black, we plot the envelope obtained by solving Eq.~\eqref{ampeq1d}. We observe excellent agreement, indicating that the two-dimensional dynamics of Eq.~\eqref{GPEr} in the weakly nonlinear regime can indeed be captured by the one-dimensional amplitude Eq.~\eqref{ampeq1d}.

The initial symmetry breaking is shown in Fig.~\ref{fig106}, where the uniform $C = 1$ state
(see Fig.~\ref{fig1}) evolves into a spatially periodic state. This is due to a MI in Eq.~\eqref{ampeq1d}, which will be discussed in Sec.~\ref{sec:MI}. Over a  relatively shorter time scale, the envelope enters the weakly nonlinear regime, oscillating between a slightly localized state (see Fig.~\ref{fig121}) and a highly localized state
(see Fig.~\ref{fig129}).
This stage can still be accurately captured by our effective one-dimensional model.
It then enters the fully nonlinear regime (see Fig.~\ref{fig64}) as two of the localized regions become dominant and results in the formation of two vortices that then get pulled into the bulk of the condensate (see Fig.~\ref{fig68}). As seen in Figs.~\ref{fig64} and \ref{fig68}, these vortices manifest as dips in the surface of $|w|$. In this regime, the weakly nonlinear results are no longer applicable. However, it is clear that the initial MI
(see Fig.~\ref{fig106}) in the one-dimensional amplitude Eq.~\eqref{ampeq1d} is the symmetry breaking mechanism responsible for the initiation of the process leading to the formation of vortices in the two-dimensional system \eqref{GPEr}.

To show that Eq.~\eqref{ampeq1d} is equivalent to (a dissipative
variant of) the self-focusing NLSE when $\gamma = 0$, we introduce the rescaled variables
\begin{equation} \label{rescale2}
T = \tau_1(1+\widetilde{\gamma}^2) \tau, \quad
Y = \sqrt{D_{YY}} \eta, \quad
C(Y,T) = \sqrt{\frac{2}{\alpha}} B(\eta, \tau); \quad {\rm where} \quad
\widetilde{\gamma} \equiv \frac{\gamma}{\tau_1},
\end{equation}
to obtain
\begin{equation} \label{ampeqB}
\frac{\widetilde{\gamma} - i}{1+\widetilde{\gamma}^2}B_\tau = B_{\eta\eta} + \sigma B + 2|B|^2B.
\end{equation}
Lastly, we multiply Eq.~\eqref{ampeqB} across by $\widetilde{\gamma} + i$ and scale out the rotation by letting
\begin{equation} \label{rotscale}
B(\eta, \tau) = e^{i\sigma \tau} A(\eta, \tau),
\end{equation}
to obtain
\begin{equation} \label{ampeqA}
A_\tau = (\widetilde{\gamma} + i) A_{\eta\eta} + \sigma \widetilde{\gamma} A + 2(\widetilde{\gamma} + i)|A|^2A.
\end{equation}
Setting $\widetilde{\gamma} = 0$ in Eq.~\eqref{ampeqA}, we see that Eq.~\eqref{ampeq1d} is equivalent to the self-focusing nonlinear NLSE. Due to rotation ($A \to Ae^{i\theta}$) and dilation ($A \to \lambda A$, $\eta \to \lambda \eta$, and $\tau \to \lambda^2\tau$) invariance the self-focusing NLSE admits a family of one-soliton solutions of the form
$$
	A_s(\eta,\tau; v, r) = r\,\sech\left\lbrack\,r(\eta + 2v\tau)\,\right\rbrack e^{-i\theta(\eta,\tau)}; \qquad \theta(\eta,\tau) = v \eta + (v^2 - r^2)\,\tau.
$$
For $\widetilde{\gamma} \ll 1$ in Eq.~\eqref{ampeqA}, a perturbation analysis invoking
additional translation ($\eta \to \eta_0$) and Galilean ($A \to Ae^{ic\eta - ic^2\tau}$ and
$\eta \to \eta - 2c\tau$) symmetries leads to a coupled system of equations for the slow time evolution of $r(\widetilde{\gamma}\tau)$ and $v(\widetilde{\gamma}\tau)$
(see, e.g.,
Refs.~\cite{elphick1990comment, fauve1990solitary, elphick1989localized, kivshar1989dynamics} for details).

Based on the time scales in Figs.~\ref{envelope}, we find that, once vortices form, they quickly enter the fully nonlinear regime. As such, a detailed analysis of the evolution of a localized soliton solution in the amplitude equation, the latter of which is only valid in the weakly nonlinear regime of Eq.~\eqref{GPEr}, is not particularly useful.
We are presently not aware of a technique (aside from the detailed numerical
simulations, such as those of Fig.~\ref{fig64}--\ref{fig68}, that
could capture this second stage of (large amplitude) symmetry breaking.
Instead, we focus on the initial symmetry breaking mechanism in Eq.~\eqref{ampeq1d} that initiates the formation of vortices in Eq.~\eqref{GPEr}. As seen in Fig.~\ref{fig106}, this symmetry breaking does occur in the weakly nonlinear regime of Eq.~\eqref{GPEr}, and is the result of a MI in Eq.~\eqref{ampeq1d}. We analyze this instability in the following section.

\section{Modulational instability}
\label{sec:MI}

In this section we analyze the MI of a spatially homogeneous time-dependent solution of Eq.~\eqref{ampeq1d}. The analysis follows that of Ref.~\cite{rapti2004modulational}; see also Ref.~\cite{segur}. To obtain an exact solution of Eq.~\eqref{ampeqA} without the spatial term, we take the ansatz 
$$A = A_0(\tau) = f(\tau)e^{ig(\tau)},$$
where the functions $f$ and $g$ satisfy the ODE's
\begin{equation} \label{fg}
\begin{gathered}
f^\prime = \widetilde{\gamma}\lbrack \sigma f + 2f^3\rbrack, \qquad f(0) = |A(0)|;
\\[2.0ex]
g^\prime = 2f^2, \qquad g(0) = \arg(A(0)).
\end{gathered}
\end{equation}
The system \eqref{fg} is solved analytically, yielding
\begin{equation}
\label{cwsolutionall}
 f(\tau) = \frac{\sqrt{\sigma}}{\sqrt{-2 + c_1 e^{-2\sigma\widetilde{\gamma}\tau}}}; \quad
 g(\tau) = -\frac{1}{2\widetilde{\gamma}} \log\left\lbrack -2 + c_1e^{-2\sigma\widetilde{\gamma}\tau}\right\rbrack - \sigma\tau + c_2,
\end{equation}
where
$$
 c_1 \equiv 2 + \frac{\sigma}{|A(0)|^2}; \qquad
 c_2 \equiv \frac{1}{2\gamma}\log\left(\frac{\sigma}{|A(0)|^2} \right) + \arg(A(0)).
$$
A spatially homogeneous solution $C_0(T)$ of Eq.~\eqref{ampeq1d} is then given by Eq.~\eqref{cwsolutionall} and the scalings in Eqs.~\eqref{rescale2} and \eqref{rotscale}. In particular, we calculate that
\begin{equation} \label{absCsq}
|C_0|^2 = \frac{2\sigma}{\alpha} \frac{1}{-2 + c_0 e^{-\frac{2\widetilde{\gamma}}{1+\widetilde{\gamma}^2}\frac{\sigma}{\tau_1}T}}; \qquad
 c_0 \equiv 2 + \frac{2\sigma}{\alpha |C_0(0)|^2}.
\end{equation}
In what follows, we let $C \to u$, $T \to t$, $Y \to y$, and $D_{YY} \to D$ for cleaner notation.

To analyze the stability of $u_0(t)$, we introduce the perturbation
\begin{equation} \label{cwpert}
	u(y,t) = u_0(t)(1 + \varepsilon w(t) \cos qy); \qquad \varepsilon \ll 1.
\end{equation}
Substituting Eq.~\eqref{cwpert} into Eq.~\eqref{ampeq1d} and equating coefficients
of $\cos qy$ at the leading order in $\varepsilon$, we obtain
\begin{equation} \label{cwlin}
	(i\tau_1 - \gamma)\left\lbrack \frac{u_0^\prime}{u_0} w + w^\prime\right\rbrack - q^2Dw + \sigma  w + \alpha|u_0|^2\lbrack \bar{w} + 2w\rbrack = 0.
\end{equation}
Next, noting that $u_0(t)$ is a solution of Eq.~\eqref{ampeq1d}, yields
\begin{equation} \label{id}
	(i\tau_1 - \gamma)\frac{u_0^\prime}{u_0} = - \sigma  - \alpha|u_0|^2.
\end{equation}
Substituting Eq.~\eqref{id} into Eq.~\eqref{cwlin} and simplifying yields
$$
	(i\tau_1 - \gamma)w^\prime - \left\lbrack q^2D - |u_0|^2 \right\rbrack w - \alpha |u_0|^2\bar{w} = 0.
$$
Now, letting $w = w_r + iw_i$ and separating real and imaginary parts, yields the 
matrix eigenvalue problem
\begin{equation} \label{mateig}
	\frac{d}{dt}\begin{pmatrix}
  w_r \\
  w_i
 \end{pmatrix} = \frac{1}{\gamma^2+\tau_1^2}
 \begin{pmatrix}
 	\gamma\left\lbrack (1+\alpha)|u_0|^2 - q^2D \right\rbrack & -\tau_1\left\lbrack (1-\alpha)|u_0|^2 - q^2D \right\rbrack \\[1.0ex]
 \tau_1\left\lbrack (1+\alpha)|u_0|^2 - q^2D \right\rbrack & ~~~\gamma\left\lbrack (1-\alpha)|u_0|^2 - q^2D \right\rbrack
 \end{pmatrix}
 \begin{pmatrix}
 w_r\\w_i
 \end{pmatrix}.
\end{equation}
The system \eqref{mateig} is non-autonomous due to the time-dependence of $|u_0|^2$ given in Eq.~\eqref{absCsq} (recall $C_0 \to u_0$). However, since $\gamma \ll 1$, we observe by Eq.~\eqref{fg} that $|u_0(t)|$ evolves on an asymptotically slow time scale as long as $|u_0| \ll \gamma^{-1/3}$. The ODE system \eqref{mateig} therefore takes the form $\mathbf{w}^\prime = M(\gamma t)\,\mathbf{w}$, where $M(\gamma t)$ is the two-by-two matrix in Eq.~\eqref{mateig} with entries that evolve slowly on an $\mathcal{O}(\gamma)$ time scale. This suggests a WKB ansatz for $\mathbf{w}$ of the form
\begin{equation} \label{wWKB}
  \mathbf{w} = \mathbf{v}(s) \,e^{r(s)/\gamma}; \qquad s = \gamma t.
\end{equation}
Substituting Eq.~\eqref{wWKB} into Eq.~\eqref{mateig} and collecting terms at leading order in $\gamma$, we find that $dr/ds$ satisfies the stationary eigenvalue problem $M\mathbf{v} = (dr/ds)\,\mathbf{v}$. We may thus identify $dr/ds$ with the eigenvalues of $M$ computed with its entries frozen in time. Therefore, $\mathbf{w} = (w_r, w_i)^T$ grows (decays) when the eigenvalue of $M$ with the largest real part lies on the right (left) half-plane. Scenarios in which the aforementioned eigenvalue slowly crosses from the left half-plane into the right half-plane may often lead to the phenomenon of delayed bifurcations \cite{baer1989slow, mandel1987slow}. That is, the bifurcation may not become observable until the slowly varying control parameter responsible for the eigenvalue crossing is an $\mathcal{O}(1)$ distance past the linear stability threshold. This phenomenon is absent here since, as we will observe below, the eigenvalue crossing is not slow. As such, we will say that an instability in Eq.~\eqref{mateig} has been triggered when the largest eigenvalue acquires zero real part.

The eigenvalue of the matrix in Eq.~\eqref{mateig} with largest real part is given by
\begin{equation} \label{lambda}
\lambda(q) = \frac{\textrm{tr}}{2} + \sqrt{\frac{\textrm{tr}^2}{4} - \textrm{det}}; \end{equation}
where
$$	
\textrm{tr} \equiv \frac{2\gamma\lbrack |u_0|^2 - q^2D \rbrack}{\gamma^2+\tau_1^2},
\quad{\rm and}\quad
\textrm{det} \equiv \frac{\left\lbrack (1+\alpha)|u_0|^2 - q^2D\right\rbrack\left\lbrack (1-\alpha)|u_0|^2 - q^2D\right\rbrack}{\gamma^2+\tau_1^2}.
$$
In the limit of small $\gamma$, we see that $\lambda(q)$ is $\mathcal{O}(|u_0|^2)$ and positive when $q^2$ lies in the interval $(q^2_-, q^2_+)$, with $q^2_{\pm} \equiv (1\pm \alpha)|u_0|^2/D$. This band of positively growing wavenumbers is what is responsible for the symmetry breaking mechanism that initiates the formation of vortices. To the left of this band, $\Re(\lambda(q))$ is $\mathcal{O}(\gamma)$ and positive, while to the right of this band, $\Re(\lambda(q))$ is $\mathcal{O}(\gamma q^2)$ and negative. We thus see that the presence of small $\gamma > 0$ is responsible for amplification of the low wavenumbers and dissipation of the high wavenumbers. The band of instability, and its $\mathcal{O}(|u_0|^2)$ positive growth rate, would be present even in the case of $\gamma = 0$. However, small $\gamma$ still influences pattern formation in Eq.~\eqref{ampeq1d} through the growth of $|u_0|^2$ by shifting the band of instability towards larger wavenumbers. On a finite domain of length $L$, in which the shortest admissible wavelength $q_1 = 2\pi/L$ may initially lie to the right of the band of instability, positive $\gamma$ will cause the band to drift rightwards and eventually trigger the instability when $q^2_+ = q_1^2$.
This occurs when $|u_0| = \mathcal{O}(1)$. To see why this precludes the delayed bifurcations, we observe that as soon as $\lambda_1$ acquires positive growth rate, its growth rate is $\mathcal{O}(1)$ positive. Therefore there is no slow crossing of the eigenvalue, and hence no delay.

\begin{figure}[htbp]
  \begin{center}
    \mbox{
       \hspace{-0.2cm}
    \subfigure[$\Re(\lambda)$ vs $q$] 
        {\label{dispersion_2}
        \includegraphics[width=.5\textwidth]{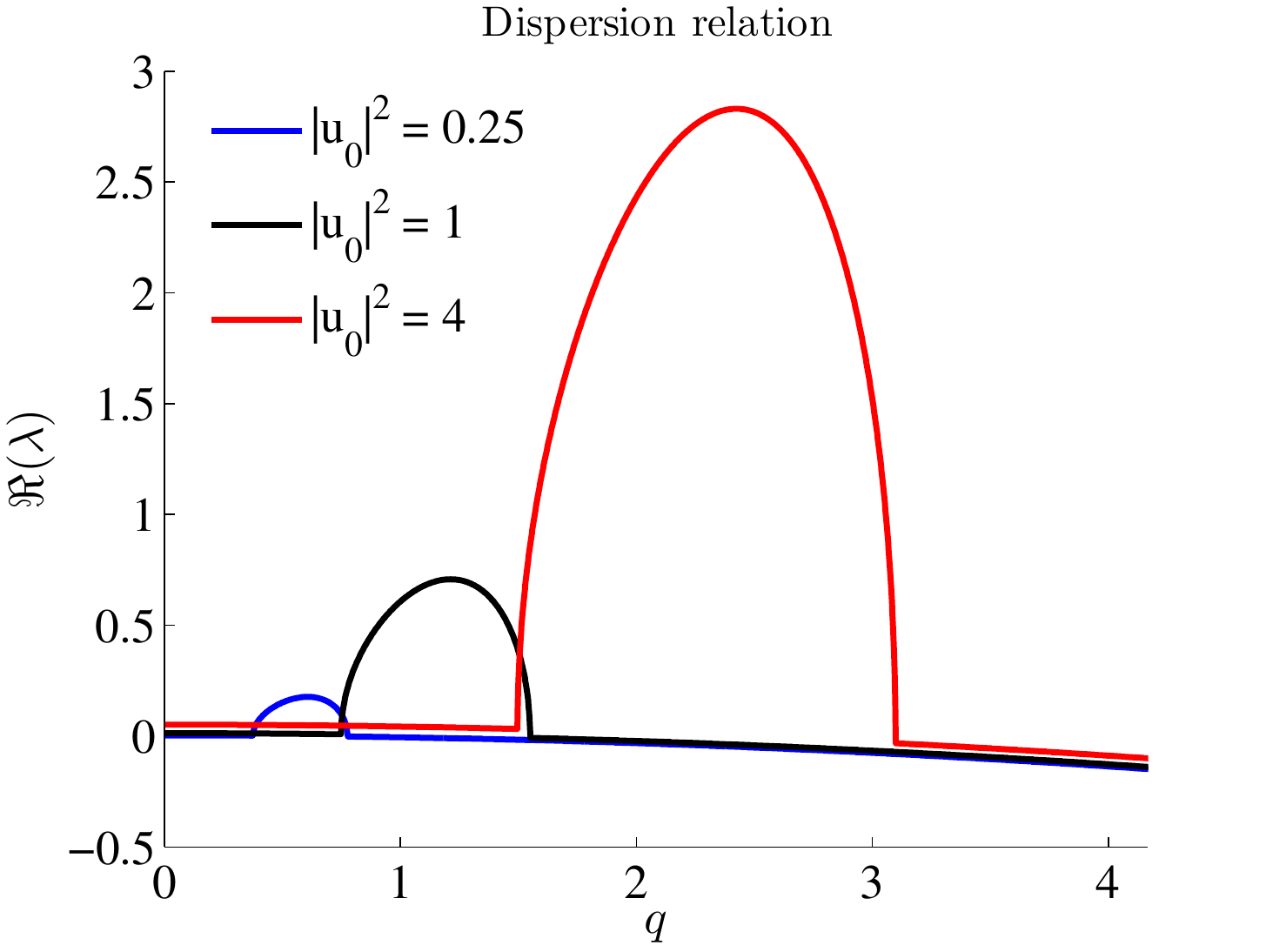}
        }   \hspace{-0.5cm}
    \subfigure[$|u_0|^2$ vs $t$] 
        {\label{u0sq}
        \includegraphics[width=.5\textwidth]{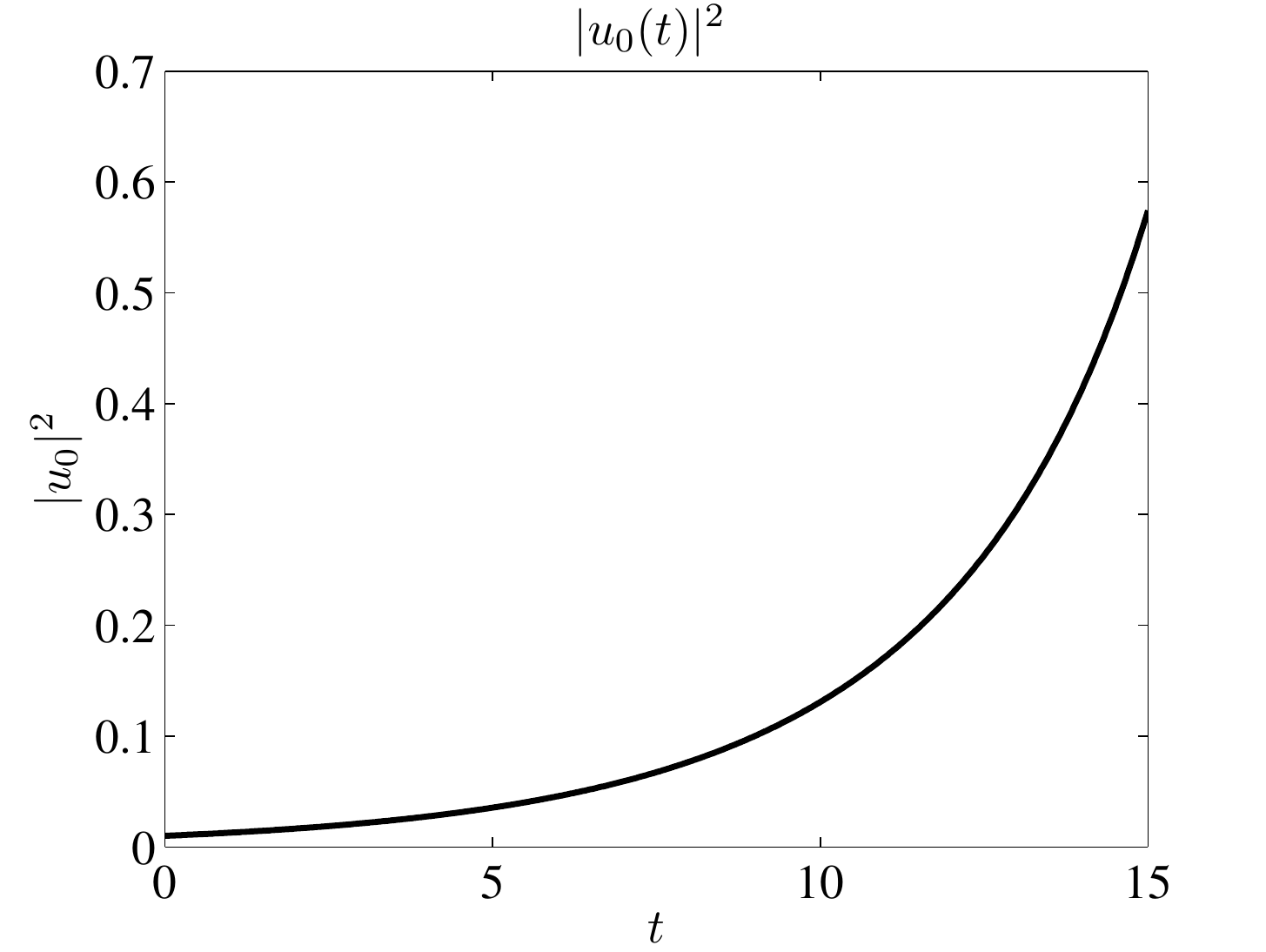}
        }}
     \caption{(a) The dispersion relation given by Eq.~\eqref{lambda} for various values of $|u_0|^2$ (0.25, 1, and 4 from left to right) as indicated in the legend. As $|u_0|^2$ increases, the band of unstable wavenumbers broadens, shifts to the right, and acquires larger growth rates. The height and shape of the bands depend only weakly on $\gamma$ when $\gamma$ is small. (b) The time evolution of $|u_0|^2$ given by Eq.~\eqref{absCsq}. The parameters are $\gamma = 0.01$ and $\Omega_2 = 1$.} \label{disp}
  \end{center}
\end{figure}

\begin{figure}[htbp]
  \begin{center}
  	\mbox{
  	\hspace{-0.2cm}
  	\subfigure[amplitude equation vs. ODE]
        {\label{compare_2}
        \includegraphics[width=.5\textwidth]{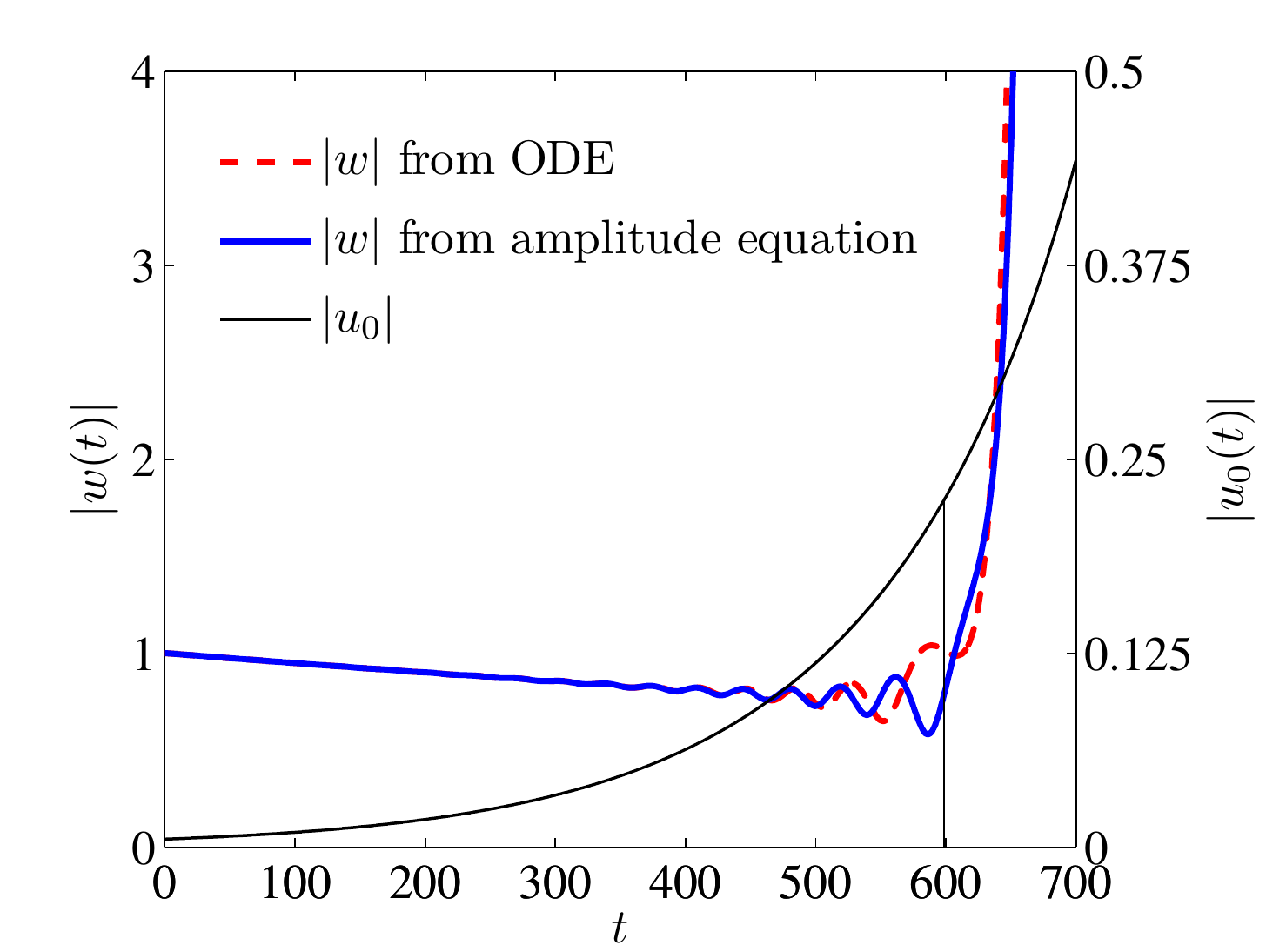}
        }\hspace{-0.3cm}
    \subfigure[full PDE system vs. amplitude equation]
    		{\label{MIfullPDE}
    		\includegraphics[width=.5\textwidth]{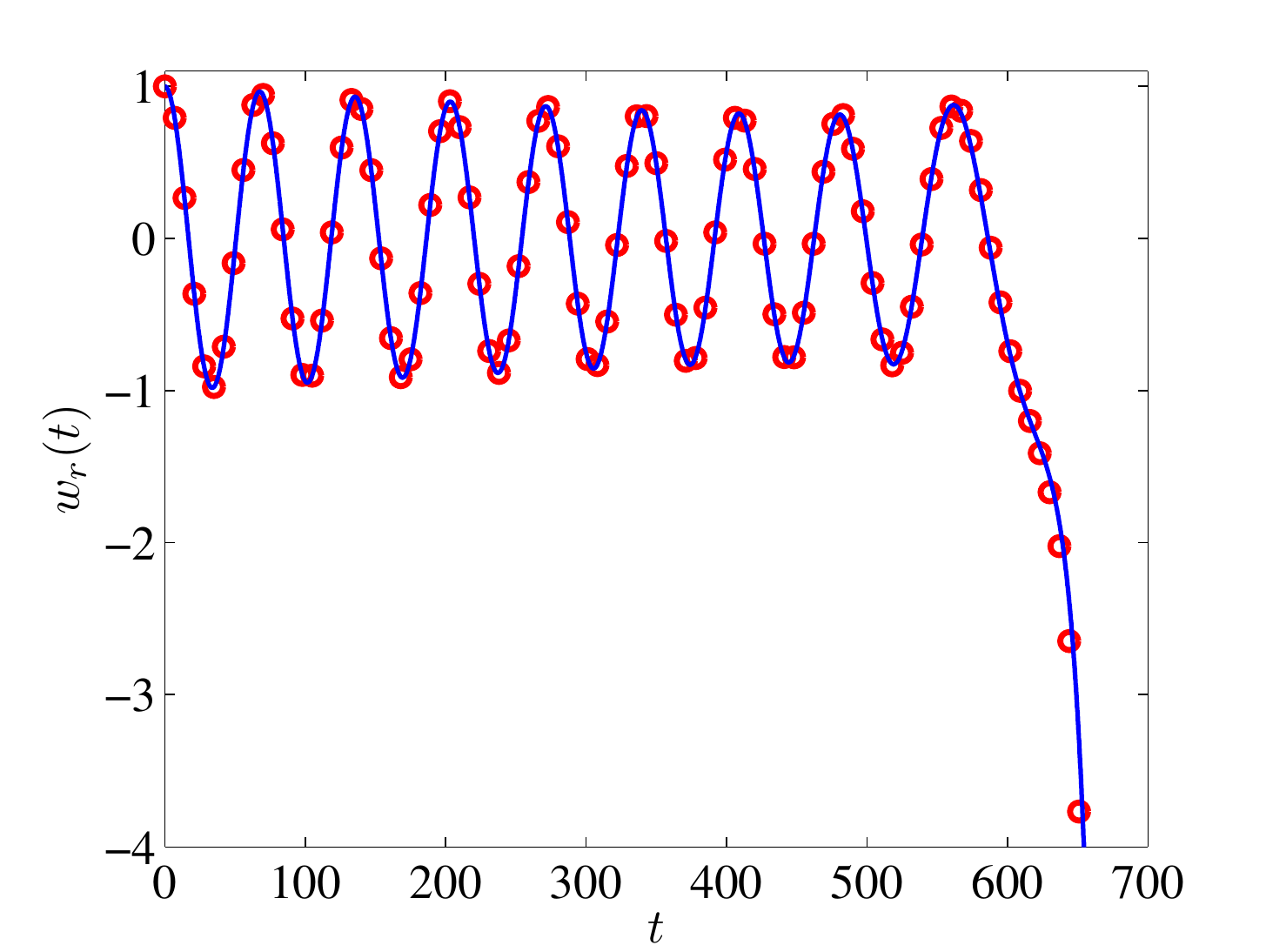}
    		}}
     \caption{(a) Growth rate of the perturbation at the periphery of the atomic cloud. The thick solid blue line (using the left axis) depicts $|w(t)|$ as numerically extracted from the PDE solution of Eq.~\eqref{ampeq1d} using the perturbation prescribed by Eq.~\eqref{cwpert}. The dashed red line depicts $|w(t)|$ computed from the linearized ODE \eqref{mateig}. The light solid line black depicts $|u_0(t)|$ (using the right axis). Both the ODE and PDE dynamics show that $|w(t)|$ decays initially until $|u_0(t)|$ has increased enough so that the smallest admissible wavenumber acquires positive growth rate. This happens approximately when $|u_0(t)|$ reaches $0.224$ (vertical line). While the ODE dynamics seems to exhibit a cumulative phase error with respect to the PDE dynamics, it does accurately predict when the instability is triggered. (b) Comparison of $\Re(w)$ as extracted from the time evolution of the full two-dimensional system \eqref{GPEr} (red circles) and that extracted from the same time evolution of the amplitude equation \eqref{ampeq1d} (blue solid) as in (a). The horizontal axis in both figures is of slow time. The parameters are $\delta = 0.04$, $\gamma = 0.005$, $\delta L_y \approx 18.09$, and $\Omega_2 = 1$.} 
  \end{center}
\end{figure}

It is important to note that the larger the $L$, where $L = \delta L_y$, the greater the number of wavelengths of the unstable mode(s) the domain can contain. Relating this back to the original system \eqref{GPEr}, the farther the rotation frequency $\Omega$ is set above threshold, the more localized regions form in the weakly nonlinear regime. This may lead to more vortices in the fully nonlinear regime being pulled into the bulk. The dependence of $\lambda(q)$ on $|u_0|^2$, along with the time evolution of $|u_0|^2$, are shown in Figs.~\ref{dispersion_2} and \ref{u0sq}, respectively.

We illustrate the theory using Fig.~\ref{envelope}, where the spatially homogeneous state being perturbed is $|u_0| = 1$. The domain length is $\delta L_y \approx 18.09$ so that, by Fig.~\ref{dispersion_2}, the third ($q \approx 1.04$) and the fourth ($q \approx 1.39$) modes are the two modes that lie in the band of instability, both having similar growth rates. Consistent with the theory, three bumps appear in Fig.~\ref{fig106}. Four bumps may also form given the same parameter set and different random initial conditions.

Lastly, we demonstrate how the growth of $|u_0|$ due to positive $\gamma$ can intrinsically trigger a MI. On a domain of length $L \approx 18.09$, we solve Eq.~\eqref{ampeq1d} with $|u_0(0)| = 0.005$ and $\gamma = 0.005$. The smallest admissible wavenumber in this domain is $q_1 = 2\pi/L \approx 0.3473$. According to Eq.~\eqref{lambda}, this wavenumber initially lies to the right of the band of instability. Therefore, the spatially homogeneous state is initially stable. However, as $|u_0|$ increases according to Eq.~\eqref{absCsq}, the band of instability drifts to the right. When $|u_0|$ increases to approximately $|u_0| \approx 0.224$, $\Re(\lambda(q_1))$ becomes positive, and the mode $\cos(q_1y)$ begins to grow. This is illustrated in Fig.~\ref{compare_2}, where we initialize $u$ as $u(y,0) = |u_0(0)|(1 + 2\times 10^{-5}\cos(q_1 y))$. The figure depicts with a thick solid blue line the evolution of $|w(t)|$ as numerically extracted from the PDE solution of Eq.~\eqref{ampeq1d} using the perturbation prescribed by Eq.~\eqref{cwpert}. The dashed red line depicts the evolution of $|w(t)|$ as computed from the linearized ODE \eqref{mateig}. The initial (oscillatory) decay is due to $q_1$ lying initially to the right of the instability band so that $\lambda(q_1)$ lies in the left half-plane with $\Im(\lambda(q_1)) \neq 0$. When $|u_0(t)|$ (light solid black line and right axis) increases to approximately $0.224$ (thin vertical line),  $\Re(\lambda(q_1))$ becomes positive real so that the amplitude of the perturbation begins to increase monotonically, verifying the theory. While the ODE prediction appears to exhibit a cumulative error in the phase with respect to the PDE dynamics, it does accurately predict when the dynamically instability is triggered. 

We next verify that this intrinsic triggering predicted by the MI analysis is also present in the full two-dimensional system \eqref{GPEr}. In Fig.~\ref{MIfullPDE}, we compare $\Re(w)$ as extracted from the time evolution of the full PDE system \eqref{GPEr} (red circles) and that extracted from the same time evolution of the amplitude equation \eqref{ampeq1d} (blue solid) as in Fig.~\ref{compare_2}. The parameters in the full PDE were taken to be $\delta = 0.04$, $\gamma = 0.005$, $\delta L_y \approx 18.09$, and $\Omega_2 = 1$. We observe excellent agreement between the two dynamics. As in Fig.~\ref{compare_2}, Fig.~\ref{MIfullPDE} exhibits a slow oscillatory decay followed by fast monotonic growth starting near $t = 600$. Note that the independent variable in the horizontal axis in both figures is the rescaled slow time $T = \delta^2 t$, which has been relabeled $t$ in accordance with the notation change $T \to t$ in this section.

To extract $\Re(w)$ from the full PDE data, we first calculate from Eq.~\eqref{O1sol} that for a given slice $x = x_0$,
\begin{equation} \label{extract1}
\frac{\psi_1(x_0,y)}{2B(x_0)} = C_r(Y,T)\cos(m_0 y) - C_i(Y,T)\sin(m_0 y),  \end{equation}
where we have defined $C_r \equiv \Re(C)$, $C_i \equiv \Im(C)$, and $Y = \delta y$ and $T = \delta^2 t$ are the slow space and time variables, respectively. Because of the separation of scales between $y$ and $Y$ in Eq.~\eqref{extract1}, the quantity $\psi_1(x_0, y)/(2B(x_0))$ to leading order takes the form of a slowly modulated phase-shifted cosine of frequency $m_0$, the envelope of which is given by $|C(Y,T)|$. Next, we calculate from Eq.~\eqref{cwpert} with $u \to C$,$t \to T$, and $y \to Y$, that
\begin{equation} \label{extract2}
\frac{1}{2\varepsilon}\frac{|C(Y,T)|^2}{|C_0(T)|^2} = \frac{1}{2\varepsilon} + \Re(w) \cos qY + \mathcal{O}(\varepsilon). \end{equation}
Here, the evolution of $|C_0(T)|^2$ is given analytically by Eq.~\eqref{absCsq}. Thus, to compute $\Re(w)$, we need only calculate numerically the envelope of the quantity $\psi_1(x_0, y)/(2B(x_0)$, which yields $|C|$. The value of $\Re(w)$ may then be extracted by numerically computing the amplitude of the left-hand side of Eq.~\eqref{extract2}, yielding the curve marked by red circles in Fig.~\ref{MIfullPDE}. This example shows that the phenomenon, predicted by the MI analysis, of initial decay of a spatial perturbation followed by growth persists not only in the reduced amplitude equation, but also in the original two-dimensional PDE system.

\section{Discussion}
\label{sec:conclu}

In the present work, we have revisited the long studied (not only theoretically
and numerically, but importantly also experimentally) problem
of the formation of vortices in the presence of rotation.
We have argued that while a vast literature exists on the subject,
there are still various gaps in our understandings of this process,
including among other things the weakly (and strongly) nonlinear
emergence of a single (or a few) vortices that eventually travel
inward, settling towards the center of the domain.
In order to shed light in the weakly nonlinear aspect within this process,
we have derived a one-dimensional effective amplitude equation as a reduction
of a dissipative variant of the self-defocusing two-dimensional GPE
with a harmonic trap under rotation. Remarkably, this equation turns out to be
a self-focusing dissipative variant of the GPE.
The latter has been shown to undergo modulational instabilities and symmetry
breakings that eventually result in the formation of solitons that lead to the
appearance of the vortices drawn inwards in the original (full)
problem. This is due to two separate symmetry breaking processes. The first, attributed to a linear (modulational) instability of a vortex-free,
homogeneous steady state of the dissipative GPE as the rotation is increased above a threshold, leads to a large number of ``small vortices'' nucleating near the edge of the condensate cloud. The second, which we can monitor numerically, but
which is beyond the realm of our weakly nonlinear theory,
selects a fraction of these small vortices and pulls them into the bulk of the condensate. 
%
%
Not only were we able to derive an effectively one-dimensional
equation describing the weakly nonlinear state (its one-dimensionality
hinting at an approximate topological insulation of the system's
boundary), but we were also able to quantify the modulational instability
and illustrate that its temporal and spatial scales coincide with
the emergence of the pattern formation within the full PDE system.
%
%
While we could not capture the final highly nonlinear step of this
destabilization and symmetry breaking process analytically, our numerical
computations shed considerable light to it. Nevertheless, the latter
would be an extremely intriguing problem for future study. While
the specific pattern selection might be the most difficult step to tackle,
it would also be interesting to perform an analysis along the lines of Ref.~\cite{weinan1994dynamics} to derive a system of equations of motion 
for the vortices as they move into the bulk.

Another key problem worth exploring, as indicated in the introduction,
is the reconciliation of the surface dynamical picture put
forth by Ref.~\cite{anglin} (see also, e.g., for a recent exposition, Ref.~\cite{dubessy}
and our earlier work of Ref.~\cite{carretero2014vortex}) and the bulk
hydrodynamic approach of Ref.~\cite{recati}.
Lastly, it would also be relevant to perform an analysis of vortex formation in the GPE with an anisotropic potential \cite{mcendoo2009small}. In the isotropic case considered here, the initial instability leads to a uniform formation of small vortices all around the edge of the condensate cloud. In contrast, in the anisotropic case, this uniformity is expected to be broken. An analysis could be performed to determine where the first vortices are nucleated, and what the subsequent vortex selection mechanism is. These problems are currently under study and relevant
progress will be reported in future publications.

\section*{Acknowledgments}

J.C.T.~was supported by an AARMS Postdoctoral Fellowship.
P.G. Kevrekidis was supported by
NSF-DMS-1312856, as well as from
the US-AFOSR under grant FA950-12-1-0332,
and the ERC under FP7, Marie Curie Actions, People,
International Research Staff Exchange Scheme (IRSES-605096).
P.G.K.'s work at Los Alamos is supported in part by the U.S. Department of Energy.
T.K.~was supported by NSERC Discovery Grant No.~RGPIN-33798 and
Accelerator Supplement Grant No.~RGPAS/461907.
R.C.G.~gratefully acknowledges the support of NSF-DMS-1309035.

\bibliographystyle{siam}
\bibliography{wnabib2}

\end{document}